\documentclass[ALICE,manyauthors]{cernphprep}
\usepackage[comma,square,numbers,sort&compress]{natbib}
\usepackage{hyperref}
\usepackage{lineno}
\usepackage{xspace}
\usepackage{xcolor}
\usepackage[T1]{fontenc}
\begin{document}
%

\newcommand{\pp}           {pp\xspace}
\newcommand{\ppbar}        {\mbox{$\mathrm {p\overline{p}}$}\xspace}
\newcommand{\XeXe}         {\mbox{Xe--Xe}\xspace}
\newcommand{\PbPb}         {\mbox{Pb--Pb}\xspace}
\newcommand{\pA}           {\mbox{pA}\xspace}
\newcommand{\pPb}          {\mbox{p--Pb}\xspace}
\newcommand{\AuAu}         {\mbox{Au--Au}\xspace}
\newcommand{\dAu}          {\mbox{d--Au}\xspace}

\newcommand{\s}            {\ensuremath{\sqrt{s}}\xspace}
\newcommand{\snn}          {\ensuremath{\sqrt{s_{\mathrm{NN}}}}\xspace}
\newcommand{\pt}           {\ensuremath{p_{\rm T}}\xspace}
\newcommand{\kt}           {\ensuremath{k_{\rm T}}\xspace} 
\newcommand{\meanpt}       {$\langle p_{\mathrm{T}}\rangle$\xspace}
\newcommand{\minv}         {\ensuremath{m_{\rm inv.}}\xspace}
\newcommand{\mee}          {\ensuremath{m_{\rm ee}}\xspace}
\newcommand{\ptee}         {\ensuremath{p_{\rm T}^{\rm ee}}\xspace}
\newcommand{\ycms}         {\ensuremath{y_{\rm CMS}}\xspace}
\newcommand{\ylab}         {\ensuremath{y_{\rm lab}}\xspace}
\newcommand{\etarange}[1]  {\mbox{$\left | \eta \right |~<~#1$}}
\newcommand{\yrange}[1]    {\mbox{$\left | y \right |~<~#1$}}
\newcommand{\dndy}         {\ensuremath{\mathrm{d}N_\mathrm{ch}/\mathrm{d}y}\xspace}
\newcommand{\dndeta}       {\ensuremath{\mathrm{d}N_\mathrm{ch}/\mathrm{d}\eta}\xspace}
\newcommand{\avdndeta}     {\ensuremath{\langle\dndeta\rangle}\xspace}
\newcommand{\dNdy}         {\ensuremath{\mathrm{d}N_\mathrm{ch}/\mathrm{d}y}\xspace}
\newcommand{\Npart}        {\ensuremath{N_\mathrm{part}}\xspace}
\newcommand{\Ncoll}        {\ensuremath{N_\mathrm{coll}}\xspace}
\newcommand{\dEdx}         {\ensuremath{\textrm{d}E/\textrm{d}x}\xspace}
\newcommand{\RpPb}         {\ensuremath{R_{\rm pPb}}\xspace}
\newcommand{\mpt}          {\ensuremath{\langle p_{\rm T}\rangle}\xspace}
\newcommand{\mptsquared}   {\ensuremath{\langle p^2_{\rm T}\rangle}\xspace}

\newcommand{\nineH}        {$\sqrt{s}~=~0.9$~Te\kern-.1emV\xspace}
\newcommand{\seven}        {$\sqrt{s}~=~7$~Te\kern-.1emV\xspace}
\newcommand{\twoH}         {$\sqrt{s}~=~0.2$~Te\kern-.1emV\xspace}
\newcommand{\twosevensix}  {$\sqrt{s}~=~2.76$~Te\kern-.1emV\xspace}
\newcommand{\five}         {$\sqrt{s}~=~5.02$~Te\kern-.1emV\xspace}
\newcommand{\thirteen}     {$\sqrt{s}~=~13$~Te\kern-.1emV\xspace}
\newcommand{\twosevensixnn}{$\sqrt{s_{\mathrm{NN}}}~=~2.76$~Te\kern-.1emV\xspace}
\newcommand{\fivenn}       {$\sqrt{s_{\mathrm{NN}}}~=~5.02$~Te\kern-.1emV\xspace}
\newcommand{\LT}           {L{\'e}vy-Tsallis\xspace}
\newcommand{\GeVc}         {Ge\kern-.1emV/$c$\xspace}
\newcommand{\MeVc}         {Me\kern-.1emV/$c$\xspace}
\newcommand{\TeV}          {Te\kern-.1emV\xspace}
\newcommand{\GeV}          {Ge\kern-.1emV\xspace}
\newcommand{\MeV}          {Me\kern-.1emV\xspace}
\newcommand{\GeVmass}      {Ge\kern-.2emV/$c^2$\xspace}
\newcommand{\MeVmass}      {Me\kern-.2emV/$c^2$\xspace}
\newcommand{\lumi}         {\ensuremath{\mathcal{L}}\xspace}

\newcommand{\ITS}          {\rm{ITS}\xspace}
\newcommand{\TOF}          {\rm{TOF}\xspace}
\newcommand{\ZDC}          {\rm{ZDC}\xspace}
\newcommand{\ZDCs}         {\rm{ZDCs}\xspace}
\newcommand{\ZNA}          {\rm{ZNA}\xspace}
\newcommand{\ZNC}          {\rm{ZNC}\xspace}
\newcommand{\SPD}          {\rm{SPD}\xspace}
\newcommand{\SDD}          {\rm{SDD}\xspace}
\newcommand{\SSD}          {\rm{SSD}\xspace}
\newcommand{\TPC}          {\rm{TPC}\xspace}
\newcommand{\TRD}          {\rm{TRD}\xspace}
\newcommand{\VZERO}        {\rm{V0}\xspace}
\newcommand{\VZEROA}       {\rm{V0A}\xspace}
\newcommand{\VZEROC}       {\rm{V0C}\xspace}
\newcommand{\Vdecay} 	   {\ensuremath{V^{0}}\xspace}

\newcommand{\jpsi}         {\ensuremath{\text{J}/\psi}\xspace}
\newcommand{\psiprime}     {\ensuremath{\psi(2\text{S})}\xspace}
\newcommand{\ccbar}        {\ensuremath{\text{c}\overline{\text{c}}}\xspace}
\newcommand{\bbbar}        {\ensuremath{\text{b}\overline{\text{b}}}\xspace}
\newcommand{\ee}           {\ensuremath{\text{e}^{+}\text{e}^{-}}\xspace} 
\newcommand{\pip}          {\ensuremath{\pi^{+}}\xspace}
\newcommand{\pim}          {\ensuremath{\pi^{-}}\xspace}
\newcommand{\kap}          {\ensuremath{\rm{K}^{+}}\xspace}
\newcommand{\kam}          {\ensuremath{\rm{K}^{-}}\xspace}
\newcommand{\pbar}         {\ensuremath{\rm\overline{p}}\xspace}
\newcommand{\kzero}        {\ensuremath{{\rm K}^{0}_{\rm{S}}}\xspace}
\newcommand{\lmb}          {\ensuremath{\Lambda}\xspace}
\newcommand{\almb}         {\ensuremath{\overline{\Lambda}}\xspace}
\newcommand{\Om}           {\ensuremath{\Omega^-}\xspace}
\newcommand{\Mo}           {\ensuremath{\overline{\Omega}^+}\xspace}
\newcommand{\X}            {\ensuremath{\Xi^-}\xspace}
\newcommand{\Ix}           {\ensuremath{\overline{\Xi}^+}\xspace}
\newcommand{\Xis}          {\ensuremath{\Xi^{\pm}}\xspace}
\newcommand{\Oms}          {\ensuremath{\Omega^{\pm}}\xspace}
\newcommand{\degree}       {\ensuremath{^{\rm o}}\xspace}

\newcommand{\note}[1]{{\color{red}\textbf{#1}}}

\begin{titlepage}
\PHyear{2021}       
\PHnumber{160}      
\PHdate{2 August}  

\title{Inclusive \jpsi production at midrapidity in pp collisions at  $\sqrt{\textbf{\textit{s}}}~=~13$~\TeV}
\ShortTitle{Inclusive \jpsi production at midrapidity in pp collisions at \thirteen}   

\Collaboration{ALICE collaboration\thanks{See Appendix~\ref{app:collab} for the list of collaboration members}}
\ShortAuthor{ALICE collaboration} 

\begin{abstract}
	We report on the inclusive \jpsi production cross section measured at the CERN Large Hadron Collider in proton--proton collisions at a centre-of-mass energy \thirteen. The \jpsi mesons are reconstructed in the \ee decay channel and the measurements are performed at midrapidity ($|y|<0.9$) in the transverse-momentum interval $0<\pt<40$~\GeVc, using a minimum-bias data sample corresponding to an integrated luminosity $L_{\text{int}} = 32.2~\text{nb}^{-1}$ and an Electromagnetic Calorimeter triggered data sample with $L_{\text{int}} =  8.3~\mathrm{pb}^{-1}$. The \pt-integrated \jpsi production cross section at midrapidity, computed using the minimum-bias data sample, is $\text{d}\sigma/\text{d}y|_{y=0} = 8.97\pm0.24~(\text{stat})\pm0.48~(\text{syst})\pm0.15~(\text{lumi})~\mu\text{b}$. An approximate logarithmic dependence with the collision energy is suggested by these results and available world data, in agreement with model predictions. The integrated and \pt-differential measurements are compared with measurements in \pp collisions at lower energies and with several recent phenomenological calculations based on the non-relativistic QCD and Color Evaporation models.
\end{abstract}
\end{titlepage}

\setcounter{page}{2} 


\section{Introduction}
\label{sec:introduction}

Quarkonium production in hadronic interactions is an excellent case of study for understanding hadronization in quantum chromodynamics (QCD), the theory of strong interactions~\cite{Brambilla:2010cs}. In particular, the production of the \jpsi meson, a bound state of a charm and an anti-charm quark and the lightest vector charmonium state, is the subject of many theoretical calculations. The cornerstone of all the theoretical approaches is the factorization theorem, according to which the \jpsi production cross section can be factorized into a short distance part describing the \ccbar production and a long distance part describing the subsequent formation of the bound state.  In this way, the \ccbar pair production cross section can be computed perturbatively. The widely used Non-Relativistic QCD (NRQCD) approach~\cite{Bodwin:1994jh} describes the transition probabilities of the pre-resonant \ccbar pairs to bound states with a set of long-distance matrix elements (LDME) fitted to experimental data, assumed to be universal. Next-to-leading order (NLO) calculations involving collinear parton densities are able to describe production yields for transverse momentum (\pt) larger than the mass of the bound state~\cite{Butenschoen:2010rq,Ma:2010yw}, but have difficulties describing the measured polarization~\cite{Chao:2012iv,Butenschoen:2012px}. Calculations employing the \kt-factorization approach~\cite{Baranov:2002cf} can reach lower \pt but have similar difficulties when compared to data~\cite{Baranov:2019lhm}. 
The low-\pt range of quarkonium production is modelled also within the Color Glass Condensate effective theory coupled to leading order NRQCD calculations~\cite{Kang:2013hta}, which involves a saturation of the small Bjorken-$x$ gluon densities that dampens the heavy-quark pair production yields.
An alternative to the universal LDME approach to hadronization used in the NRQCD framework is provided by the Color Evaporation Model (CEM)~\cite{Barger:1979js,Gavai:1994in} and its more recent implementation using the \kt-factorization approach, the Improved CEM (ICEM)~\cite{Ma:2016exq}. In the ICEM, the transition probability to a given bound state is proportional to the \ccbar pair production cross section integrated over an invariant-mass range spanning between the mass of the bound state and twice the mass of the lightest charmed meson. 
Finally, in the Color Singlet Model (CSM)~\cite{Chang:1979nn,Baier:1981uk,Berger:1980ni}, the pre-resonant \ccbar pair is produced directly in the color-singlet state with the same quantum numbers as the bound state. Calculations within this model at NLO precision are known to strongly underpredict the measured production cross sections~\cite{Lansberg:2010vq}. In this context, a \pt-differential measurement of \jpsi production cross section covering a wide \pt range, starting from \pt=0 and up to high-\pt, can discriminate between the different models of quarkonium production.

In this paper, we present the integrated, and the \pt and rapidity ($y$) differential production cross sections of inclusive \jpsi production at midrapidity ($|y|<0.9$) in proton--proton (\pp) collisions at the center-of-mass energy \thirteen. 
The inclusive \jpsi yields include contributions from directly produced \jpsi, feed-down from prompt decays of higher-mass charmonium states, and non-prompt \jpsi from the decays of beauty hadrons. The \pt-differential production cross section of inclusive \jpsi is measured in the $0<\pt<15$~\GeVc interval using a minimum-bias triggered data sample and in the $15<\pt<40$~\GeVc interval using an Electromagnetic Calorimeter triggered data sample.
These results complement existing measurements at midrapidity at \thirteen performed by the CMS Collaboration~\cite{Sirunyan:2017qdw}, which report the prompt \jpsi production cross section for \pt $>20$~\GeVc. Previous measurements of the \jpsi production cross section in \pp collisions performed by the ALICE Collaboration at midrapidity at lower energies were published in Refs.~\cite{Aamodt:2011gj,Abelev:2012kr,Acharya:2019lkw}. The inclusive \jpsi production cross section in \pp collisions at \thirteen was published by the ALICE Collaboration at forward rapidity in Ref.~\cite{Acharya:2017hjh} and the prompt and non-prompt production cross sections were reported by the LHCb Collaboration in Ref.~\cite{Aaij:2015rla}.

In the next sections, the ALICE detector and the data sample are described in Section~\ref{sec:setup}, and the data analysis and the determination of the systematic uncertainties are described in Section~\ref{sec:analysis} and Section~\ref{sec:uncertainties}, respectively. The results are presented and discussed together with recent model calculations in Section~\ref{sec:results} and conclusions are drawn in Section~\ref{sec:conclusion}.

\section{The ALICE detector, data set and event selection}
\label{sec:setup}

A detailed description of the ALICE detector and its performance is provided in Refs.~\cite{Aamodt:2008zz, Abelev:2014ffa}. Here we mention only the detector systems used for the reconstruction of the \jpsi mesons decaying in the \ee channel at midrapidity. Unless otherwise specified, the term electrons will be used throughout the text to refer to both electrons and positrons.

The reconstruction of charged-particle tracks is performed using the Inner Tracking System (ITS)~\cite{Aamodt:2010aa} and the Time Projection Chamber (TPC)~\cite{Alme:2010ke}, which are placed inside a solenoidal magnet providing a uniform magnetic field of $B = 0.5~\text{T}$ oriented along the beam direction. The ITS is a silicon detector consisting of six cylindrical layers surrounding the beam pipe at radii between 3.9 and $43.0~\text{cm}$. The two innermost layers consist of silicon pixel detectors (SPD), followed by two layers of silicon drift (SDD) and two layers of silicon strip (SSD) detectors. The TPC is a cylindrical gas drift chamber which extends radially between 85 and $250~\text{cm}$ and longitudinally over $250~\text{cm}$ on each side of the nominal interaction point. Both TPC and ITS have full coverage in azimuth and provide tracking in the pseudorapidity range $|\eta| < 0.9$. Additionally, the measurement of the specific energy loss (\dEdx) in the TPC active gas volume is used for electron identification.
The Electromagnetic Calorimeter (EMCal) and the Di-jet Calorimeter
(DCal)~\cite{Abeysekara:2010ze, Cortese:2008zza, Allen:2010stl} are employed for triggering and electron identification.
The EMCal/DCal is a Shashlik-type lead-scintillator sampling calorimeter located at a radius of 4.5~m from the beam vacuum tube.
The EMCal detector covers a pseudorapidity range of  $|\eta|<0.7$ over an azimuthal angle of $ 80^\circ < \varphi < 187^\circ$, and the DCal covers $0.22 <|\eta|< 0.7$ for $260^\circ < \varphi < 320^\circ$ and $|\eta|<0.7$ for $320^\circ < \varphi < 327^\circ$. The EMCal and DCal have identical granularity and intrinsic energy resolution, and they form a two-arm electromagnetic calorimeter, which in this paper will be referred to jointly as EMCal.

In addition to these central barrel detectors, the V0 detectors, composed of two scintillator arrays~\cite{Abbas:2013taa} placed along the beam line on either side of the interaction point and covering the pseudorapidity intervals $-3.7 < \eta < -1.7$ and $2.8 < \eta < 5.1$, respectively, are used for event triggering. Together with the SPD detector, the V0 is also used to reject background from beam-gas collisions and pileup events.

The measurement of the \pt-integrated and \pt-differential production cross sections up to $\pt = 15$~\GeVc, the upper limit being determined by the available integrated luminosity, utilizes the minimum-bias (MB) trigger, defined as the coincidence of signals in both V0 scintillator arrays. 
For the \pt interval from 15 up to 40~\GeVc, the EMCal trigger is employed to select events with high-\pt electrons. The lower \pt limit is chosen such that the trigger efficiency does not vary with \pt above this value, thus avoiding systematic uncertainties related to trigger threshold effects. 
The EMCal trigger is an online trigger which includes a Level 0 (L0) and a Level 1 (L1) component~\cite{Abeysekara:2010ze}. 
The calorimeter is segmented into towers and Trigger Region Units (TRUs), the latter being composed of 384 towers each~\cite{Abeysekara:2010ze}.
The L0 trigger is based on the analog charge sum of $4\times4$ groups of adjacent towers evaluated within each TRU, in
coincidence with the MB trigger.
The L1 trigger decision requires the L0 trigger and, in addition, scans for $4\times4$ groups of adjacent towers across the entire EMCal surface.
The EMCal-triggered analysis presented in this paper uses the L1 trigger, and requires that at least one of the charge sums of the $4\times4$ adjacent towers  is above 9~GeV.

This analysis includes all the data recorded by the ALICE Collaboration during the LHC Run 2 data-taking campaigns of 2016, 2017 and 2018 for \pp collisions at \thirteen. The maximum interaction rate for the dataset was $260~\text{kHz}$, with a maximum pileup probability in the same bunch crossing of 0.5$\times$10$^{-3}$. The events selected for analysis were required to have a reconstructed vertex within the interval $|z_{\text{vtx}}|<10~\text{cm}$ to ensure a uniform detector acceptance. Beam-gas events and pileup collisions occurring within the readout time of the SPD were rejected offline using timing selections based on the V0 detector information. Pileup collisions occurring within the same LHC bunch crossing were rejected using offline algorithms which identify multiple vertices~\cite{Abelev:2014ffa}. The remaining fraction of pileup events surviving the selections is negligible for both the MB and EMCal data samples.

The analyzed MB sample, satisfying all the quality selections, consists of about $2\times$10$^9$ events, corresponding to an integrated luminosity $L_{\text{int}} = 32.2~\text{nb}^{-1} \pm 1.6\%~(\text{syst})$, and the EMCal-triggered sample consists of approximately $9\times$10$^7$ events which corresponds to an integrated luminosity $L_{\text{int}} = 8.3~\mathrm{pb}^{-1} \pm 2.0\%~(\text{syst})$. The integrated luminosities are obtained based on the MB trigger cross section ($\sigma_{\mathrm{MB}}$), measured in a van der Meer scan~\cite{vanderMeer:296752}, separately for each year, as described in Ref.~\cite{aliceLumi13TeVrun2}. For each of the used triggers, MB and EMCal, the integrated luminosity is obtained as
\begin{equation}
L_{\mathrm{int}} = \frac{N_{\mathrm{MB}}}{\sigma_{\mathrm{MB}}} \times \frac{ds_{\mathrm{trig}}}{ds_{\mathrm{MB}}} \times \frac{LT_{\mathrm{trig}}}{LT_{\mathrm{MB}}}
\end{equation}
where $N_{\mathrm{MB}}$ is the number of MB-triggered events in the triggered sample, $ds_{\mathrm{trig}}$ is the downscaling factor applied to the considered trigger by the ALICE trigger processor and $LT_{\mathrm{trig}}$ is the trigger live time, i.e. the fraction of time where the detector cluster~\footnote{A detector cluster is a set of ALICE detectors readout with the same set of triggers. All triggers assigned to a given cluster share the same live-time, determined by the slowest detector in the cluster.} assigned to the trigger was available for readout.

\section{\jpsi reconstruction}
\label{sec:analysis}
In this work we study the integrated, and the rapidity- and \pt-differential inclusive \jpsi production at midrapidity ($|y|<0.9$) reconstructing the \jpsi from the \ee decay channel. The MB sample analysis follows closely the one performed in \pp collisions at \five~\cite{Acharya:2019lkw}. 

\subsection{Track selection}
\label{sec:trackSelection}

Electron-track candidates are reconstructed employing the ITS and TPC. They are required to be within the acceptance of the central barrel ($|\eta| < 0.9$), and to have a minimum transverse momentum of 1~\GeVc, which suppresses the background with only a moderate \jpsi efficiency loss. The tracks are selected to have at least 2 hits in the ITS, one of which having to be in one of the SPD layers, and share at most one hit with other tracks. A minimum of 70 out of a maximum of 159 clusters are required in the TPC. 

In order to reject tracks originating from weak decays and interactions with the detector material, a selection based on the distance-of-closest approach (DCA) to the primary vertex is applied to the  tracks. For the MB analysis the tracks are required to have a minimum DCA lower than 0.2~cm in the transverse direction and 0.4~cm along the beam axis. Such tight selection criterion is used in order to improve the signal-to-background ratio and the signal significance. It was checked with Monte Carlo (MC) simulations that these requirements do not lead to efficiency loss for the non-prompt \jpsi relative to the prompt \jpsi. For the EMCal-triggered event analysis, a looser selection on the DCA to the primary vertex at 1 and 3~cm is applied to avoid rejecting non-prompt \jpsi from highly boosted beauty hadron decays.

The electrons are identified using the specific energy loss \dEdx in the TPC gas. Their \dEdx is required to be within a band of $[-2, 3]~\sigma$ relative to the expectation for electrons at the given track momentum, with $\sigma$ being the \dEdx resolution. The contamination from protons which occurs in the momentum range $p<1.5$~\GeVc and from pions for momenta above 2~\GeVc is mitigated by rejecting tracks compatible with the proton or pion hypothesis within $3\sigma$. 

For the analysis of EMCal-triggered events, both the TPC and the EMCal are used for electron identification. At least one of the \jpsi decay-electron tracks, initially identified by the TPC, is required to be matched to an EMCal cluster (a group of adjacent towers belonging to the same electromagnetic shower). In order to ensure a constant trigger efficiency on the selected events, the matched clusters are selected to have a minimal energy of 14~\GeV, a value that is significantly higher than the applied online threshold of 9~\GeV. Electrons are identified by applying a selection on the energy-to-momentum ratio of the EMCal matched track of $0.8<E/p<1.3$ and on the \dEdx in the TPC of $[-2.25, 3]~\sigma$. Due to the additional use of the EMCal for electron identification with respect to the MB based analysis, no explicit hadron rejection was used for the EMCal sample.

Secondary electrons from photon conversions, the main background source for both analyses, are rejected using the requirement of a hit in the SPD detector. This requirement rejects most of the electrons from photon conversions occurring beyond the SPD layers. An additional selection based on track pairing, as described in details in Ref.~\cite{Acharya:2019lkw}, is applied to further reject conversion electrons, especially those from photons converting in the beam pipe or in the SPD. 

\subsection{Signal extraction}
\label{sec:signalExtraction}

The number of reconstructed \jpsi mesons is extracted from the invariant mass ($\mee$)  distribution of all possible opposite-sign (OS) pairs constructed combining the selected electron tracks within the same event (SE). Besides the \jpsi signal, i.e. pairs of electrons originating from the decay of a common \jpsi mother, the invariant mass distribution contains a background with contributions from combinatorial and correlated sources. 

In the MB analysis, the combinatorial background, i.e. pairs of electrons originating from uncorrelated processes, is estimated using the event-mixing technique (ME), in which pairs are built from opposite-sign electrons belonging to different events. The mixing is done considering events from the same run (a collection of events taken during a period of time of up to a few hours) with a similar vertex position. The normalized combinatorial background distribution $B_{\rm comb}(\mee)$ is obtained as

\begin{equation}
B_{\rm comb}(\mee) = N_{\rm OS}^{\rm ME}(\mee) \times \frac{\sum_{m_{\rm i}} N_{\rm LS}^{\rm SE}(m_{\rm i})}{\sum_{m_{\rm i}} N_{\rm LS}^{\rm ME}(m_{\rm i})}
\label{eq:meNorm}
\end{equation}

where $N_{\rm LS}^{\rm SE}$, $N_{\rm OS}^{\rm ME}$ and $N_{\rm LS}^{\rm ME}$ are the number of same-event like-sign (LS), mixed-event OS and mixed-event LS pairs, respectively. Here, the mixed-event OS distribution is normalized using the ratio of SE to ME like-sign pairs since these are not expected to contain any significant correlated source. The summation extends over all the mass bins $m_{\rm i}$ between 0 and 5~\GeVmass to minimize the statistical uncertainty on the background matching.
The correlated background in the mass region relevant for this analysis originates mainly from semi-leptonic decays of heavy-flavor hadrons~\cite{Acharya:2018kkj}. In order to extract the number of reconstructed \jpsi, $N_{\jpsi}$, the combinatorial background-subtracted invariant mass distribution is fitted with a two-component function: one empirical function to describe the correlated background shape, which is a second order polynomial at low pair \pt ($\ptee<1$~\GeVc) and an exponential at high-\pt ($\ptee>1$~\GeVc), plus a template shape obtained from MC simulations, described in Section~\ref{sec:Corrections}, for the \jpsi signal. 

For the analysis of the EMCal-triggered event sample, due to the relatively large contribution from correlated sources at high \pt, the event mixing technique is not used. Instead, a fit of the invariant mass distribution is performed using the MC template for the signal and a third-order polynomial function to describe both the combinatorial and correlated background contributions.

\begin{figure}[h]
    \centering
    \includegraphics[width=\textwidth]{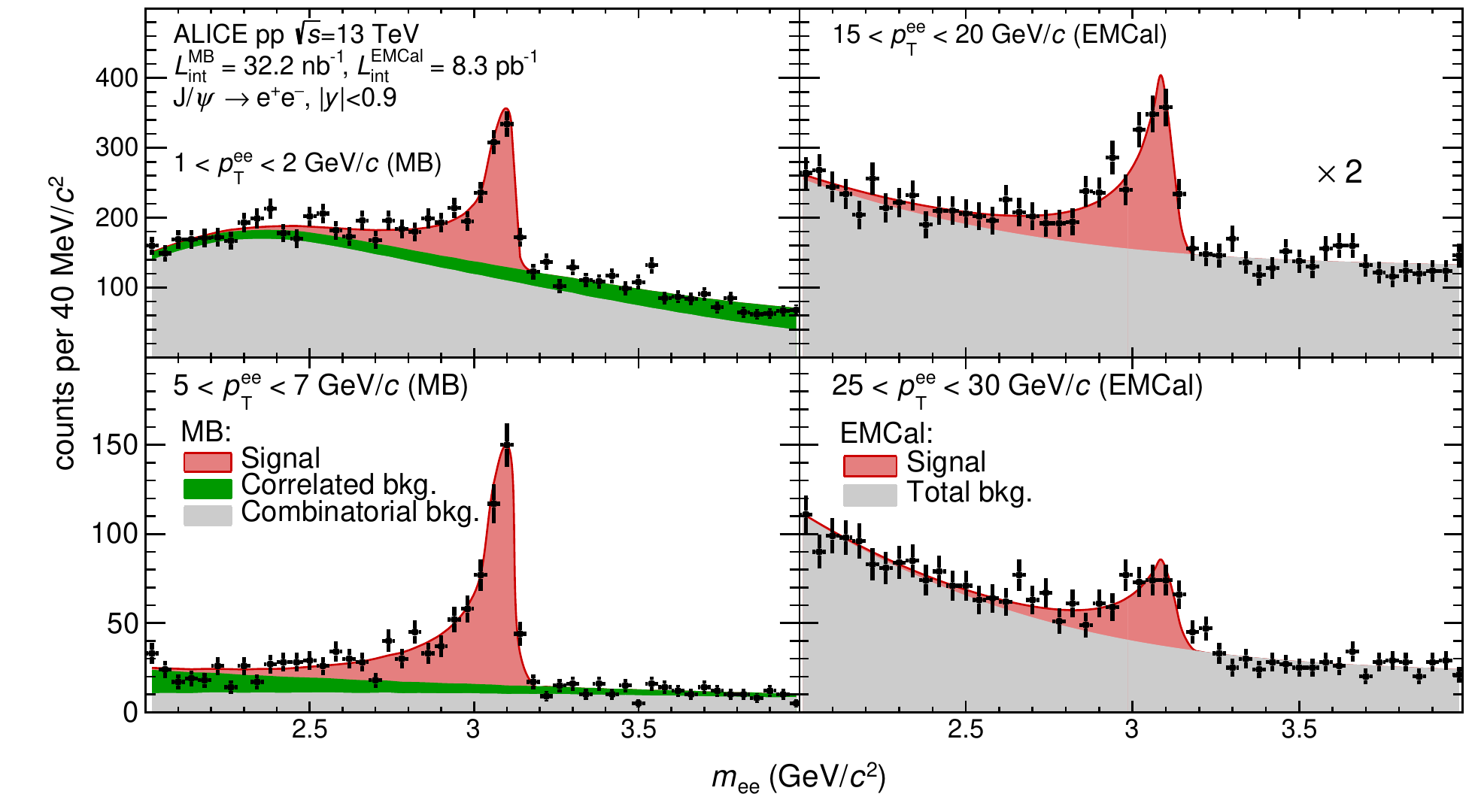}
    \caption{(Color online) Invariant-mass distributions for SE \ee pairs in two \ptee intervals from the MB event analysis (left panels) and two \ptee intervals from the EMCal-triggered event analysis (right panels). The signal and background components obtained from the fit procedure are shown separately. For the top-right panel, the distributions are scaled for convenience by a factor of 2.}
    \label{fig:signalExtraction}
\end{figure}

In both analyses, the contribution from $\psi(2S)$ decaying in the dielectron channel is not included in the fit as the expected number of such pairs is $\sim$1\% of the \jpsi raw yield and it is statistically not significant in the analyzed data samples. The number of \jpsi is obtained by counting the number of \ee pairs in the mass range $2.92 \leq m_{\text{ee}} \leq 3.16$~\GeVmass remaining after subtracting the background. The SE-OS dielectron invariant mass distribution for a few of the \pt intervals is shown in Fig.~\ref{fig:signalExtraction}, together with the estimated signal and background components. 

\subsection{Corrections}
\label{sec:Corrections}

The double differential \jpsi production cross section is calculated as

\begin{align}
    \frac{\text{d}^2\sigma_{\jpsi}}{\text{d}y\text{d}\pt} = \frac{N_{\jpsi}}{\text{BR}(\jpsi\rightarrow\ee) \times \langle A \times \epsilon \rangle \times \Delta y \times \Delta\pt \times L_{\text{int.}}} \, ,
    \label{eq:crossSection}
\end{align}

where $N_{\jpsi}$ is the number of reconstructed \jpsi in a given interval of rapidity $\Delta y$ and transverse momentum $\Delta$\pt, $\text{BR}(\jpsi\rightarrow\ee)$ is the decay branching ratio into the dielectron channel~\cite{Tanabashi:2018oca}, $\langle A \times \epsilon \rangle$ is the average acceptance and efficiency factor and $L_{\text{int}}$ is the integrated luminosity of the data sample.

The correction for acceptance and efficiency is the product of the kinematical acceptance factor, the reconstruction efficiency, which includes both tracking and particle-identification (PID) efficiency, and the fraction of signal in the signal counting mass window. For the EMCal-triggered events analysis, the efficiency for the EMCal cluster reconstruction is also considered. The efficiency related to the EMCal trigger is estimated using a parameterized simulation of the L1 trigger which, includes decalibration and noise based on measured data, and takes into account the time-dependent detector conditions. With the exception of the PID efficiency, all the corrections are obtained based on a MC simulation of unpolarized \jpsi mesons embedded in inelastic \pp collisions simulated using PYTHIA 6.4~\cite{Sjostrand:2006za} with the Perugia 2011 tune~\cite{Skands:2010ak}. The prompt \jpsi are generated with a flat rapidity distribution and a \pt spectrum obtained from a phenomenological interpolation of \jpsi measurements at RHIC, CDF and the LHC at lower energies~\cite{Bossu:2011qe}. For the non-prompt \jpsi, \bbbar pairs are generated using the PYTHIA Perugia 2011 tune. The \jpsi decays are simulated using PHOTOS~\cite{Barberio:1990ms}, which includes the radiative component of the \jpsi decay. The generated particles are transported through the ALICE detector setup using the GEANT3 package~\cite{Brun:1082634}.

The PID efficiency is determined with a data-driven method by using a clean sample of electrons from tagged photon conversion processes, passing the same quality criteria as the electrons selected for the \jpsi reconstruction. The PID selection efficiency for single electrons is propagated to the \jpsi level using a simulation of the \jpsi decay. 
The acceptance times efficiency correction factor for the MB sample analysis varies with \pt between $7.6\%$ and $16\%$ while in the case of the EMCal-triggered sample analysis it increases with \pt from 2 to 8\%.

Due to the finite size of the \pt intervals, there is a mild dependence of the correction factors on the shape of the inclusive \jpsi \pt distribution used in the simulation. This is mitigated iteratively by using the corrected \jpsi \pt-differential production cross section to reweight the acceptance times efficiency correction factor and obtain an updated corrected cross section. The procedure is stopped when the difference between the input and output corrected \pt-differential production cross section drops below 1\%, which typically occurs within 1 to 2 iterations, depending on the \pt interval.
Additionally, to check if the default MC used in the analysis could introduce a bias on the EMCal trigger efficiency, due to the enhancement of \jpsi, another MC simulation, based on a di-jet production generated by PYTHIA8~\cite{Sjostrand:2007gs}, at \thirteen, is used as a cross-check. As a result, the default MC and the di-jet MC lead to a compatible EMCal trigger efficiency.

\section{Systematic uncertainties}
\label{sec:uncertainties}

There are several sources of systematic uncertainties affecting this analysis, namely the ITS-TPC tracking, the electron identification, the signal extraction procedure, the \jpsi input kinematic distributions used in MC simulations, the determination of the integrated luminosity and the branching ratio of the dielectron decay channel. A summary of these is given in Table~\ref{tab:systematics}.

The uncertainty of the ITS-TPC tracking efficiency is one of the dominant sources of systematic uncertainty and has two contributions: one due to the TPC-ITS matching and one related to the track-quality requirements. The former is obtained from the residual difference observed for the ITS-TPC single-track matching between data and MC simulations~\cite{Acharya:2017jgo}, which is further propagated to \jpsi dielectron pairs. It varies between 2.8\% and 5.4\%, depending on \pt. The uncertainty related to the track-quality requirements is estimated by repeating the analysis with variations of the selection criteria and taking the root mean square (RMS) of the distribution of the results as systematic uncertainty. This uncertainty also depends on \pt and is equal to 3.7\% for $\pt<5$~\GeVc and approximately 2\% for $\pt>5$~\GeVc. In Table~\ref{tab:systematics}, both contributions are added in quadrature and provided as ranges for the \pt- and $y$-differential results.

As described in Section~\ref{sec:Corrections}, the particle identification efficiency is determined via a data-driven procedure using a sample of identified electrons from tagged photon conversions. For the MB data sample, the uncertainty of this procedure is estimated by repeating the analysis with a looser and a tighter hadron (pion and proton combined) rejection criteria and taking the largest deviation from the results obtained with the standard PID selection divided by $\sqrt{12}$.  In addition, the statistical uncertainty of the pure electron sample used for the determination of the efficiency, which becomes non-negligible at high \pt, is propagated to the total uncertainty for the PID. The total PID systematic uncertainty is larger than 1\% only for $\pt>7$~\GeVc, reaching 4\% in the \pt interval $10<\pt<15$~\GeVc.
For the EMCal-triggered analysis, the particle identification systematic uncertainty has contributions from the electron identification in both the TPC and the EMCal. The values are estimated by varying the \dEdx range for the electron selection in the TPC, the $E/p$ selection range in the EMCal, and the minimum energy of the matched clusters. The total PID uncertainty, obtained in an analogous way as for the MB sample, increases with \pt from 2.8\% to 5.4\%.

For $p_{\rm T}$ up to 15~\GeVc, the systematic uncertainties associated to the \jpsi signal extraction procedure is dominated by the \jpsi invariant mass signal shape template used to calculate the fraction of reconstructed \jpsi mesons within the signal counting mass window. This uncertainty amounts to 1.9\% and is evaluated by repeating the extraction of the corrected yield with different invariant mass intervals used for the signal counting and taking the RMS of the variations as a systematic uncertainty. An additional source of uncertainty, due to the fit of the correlated background, is determined by varying the fit mass range and is typically below 1\%. For $p_{\rm T} > $15~\GeVc, the systematic uncertainty associated to the \jpsi signal extraction are evaluated similarly to the MB analysis, however, the components associated to the signal shape and the background fit have similarly large contributions. This is one of the main sources of uncertainty in this \pt range. The total uncertainty of the signal extraction varies with \pt between about 2 and 7\%. 

The uncertainty on the EMCal trigger efficiency is studied varying the contribution from random noise (evaluated using energy resolution measurements in data) applied to the $4\times4$ groups of adjacent towers. It was also studied using the di-jet production generated by PYTHIA8 at \thirteen mentioned in Sec.~\ref{sec:Corrections}. The latter study showed a difference in the EMCal trigger efficiency of 0.5\%, which is assigned as a systematic uncertainty.

Since the \jpsi efficiency has a dependence on \pt, the particular \jpsi kinematic distribution used in the MC simulation, from which efficiencies are derived, can have an impact on the average efficiency computed for finite \pt intervals. As described in Section~\ref{sec:Corrections}, this is mitigated via an iterative procedure and the remaining uncertainty is related to the precision of the measured \jpsi spectrum. This uncertainty is estimated by fitting the measured \pt spectrum with a power law function and allowing the fitted parameters to vary according to the covariance matrix. For each of such a variation, the average efficiency in a given kinematic interval is recomputed and the RMS of the distribution obtained from the variation of efficiency with respect to the central value is taken as a systematic uncertainty. This amounts to 1\% for the \pt-integrated case and is smaller for all of the considered \pt intervals.

The uncertainty of the integrated luminosity arising from the vdM-scan based measurements amounts to $1.6\%$ for both the MB and EMCal-triggered data samples, and is determined as described in Ref.~\cite{aliceLumi13TeVrun2}. For the EMCal-triggered data sample only, an additional uncertainty of 1.1\% arising from the precision of the trigger downscaling is assigned.
The uncertainty on the \jpsi decay branching ratio to the dielectron channel amounts to 0.53\% as reported by the Particle Data Group~\cite{Tanabashi:2018oca}.

The uncertainties of the integrated luminosity and branching ratio are treated as global systematic uncertainties. All the other uncertainties are considered as point to point correlated, with the exception of the one due to the background fit which is considered to be fully uncorrelated. The systematic uncertainties are thus dominated by correlated sources. The total correlated, uncorrelated and global uncertainties obtained as the sum in quadrature of the corresponding sources are given in Table~\ref{tab:systematics}. 

\begin{table}[]
    \centering
    \caption{Summary of contributions to systematic uncertainties of the measured \jpsi production cross section (in percentage).}
    \begin{tabular}{l | c | c | c | c}
    Source                & (\pt, $y$)-integrated & $y$-differential & \pt-differential (MB) & \pt-differential (EMCal)  \\
                  &  & & $ 0 < \pt < 15$ (\GeVc) & $ 15 < \pt < 40$ (\GeVc)   \\

        \hline\hline
        Tracking          & \multicolumn{2}{c|}{4.9}        & 4.7 -- 6.5 & 5.7 \\
        \hline
        PID               & \multicolumn{2}{c|}{0.6}                 & 0.0 -- 4.1 & 2.8 -- 5.4  \\
        \hline
        Signal shape      & \multicolumn{2}{c|}{1.9}        & 1.4 -- 2.2  & 1.0 -- 4.0 \\
        \hline
        Background fit    & 0.3                  & 0.3 -- 0.4      & 0.2 -- 1.2 & 1.0 -- 6.0 \\
        \hline
        MC input          & \multicolumn{2}{c|}{1.0}                  & 0.0 -- 0.9 & 0.9 \\
          \hline
       EMCal trigger        &     not used            &   not used    &  not used & 0.5 \\
        \hline
        Luminosity        & \multicolumn{3}{c|}{1.6} & 2.0             \\
        \hline
        Branching ratio   & \multicolumn{4}{c}{0.5}                              \\

        \hline \hline
        Total uncorrelated & 0.3                 & 0.3 -- 0.4      & 0.2 -- 1.2 & 1.0 -- 6.0 \\
        \hline
        Total correlated   & 5.4                  & 5.4       & 5.2 -- 7.4 & 6.5 -- 8.8 \\
        \hline
        Global             & \multicolumn{3}{c|}{1.7}  &  2.0        \\
        \hline \hline
        Total (w/o global)      & 5.4                  & 5.4        & 5.3 -- 7.5 & 7.0 -- 11.0
    \end{tabular}
    \label{tab:systematics}
\end{table}

\section{Results}
\label{sec:results}

\begin{figure}
    \centering
    \includegraphics[width=.48\textwidth]{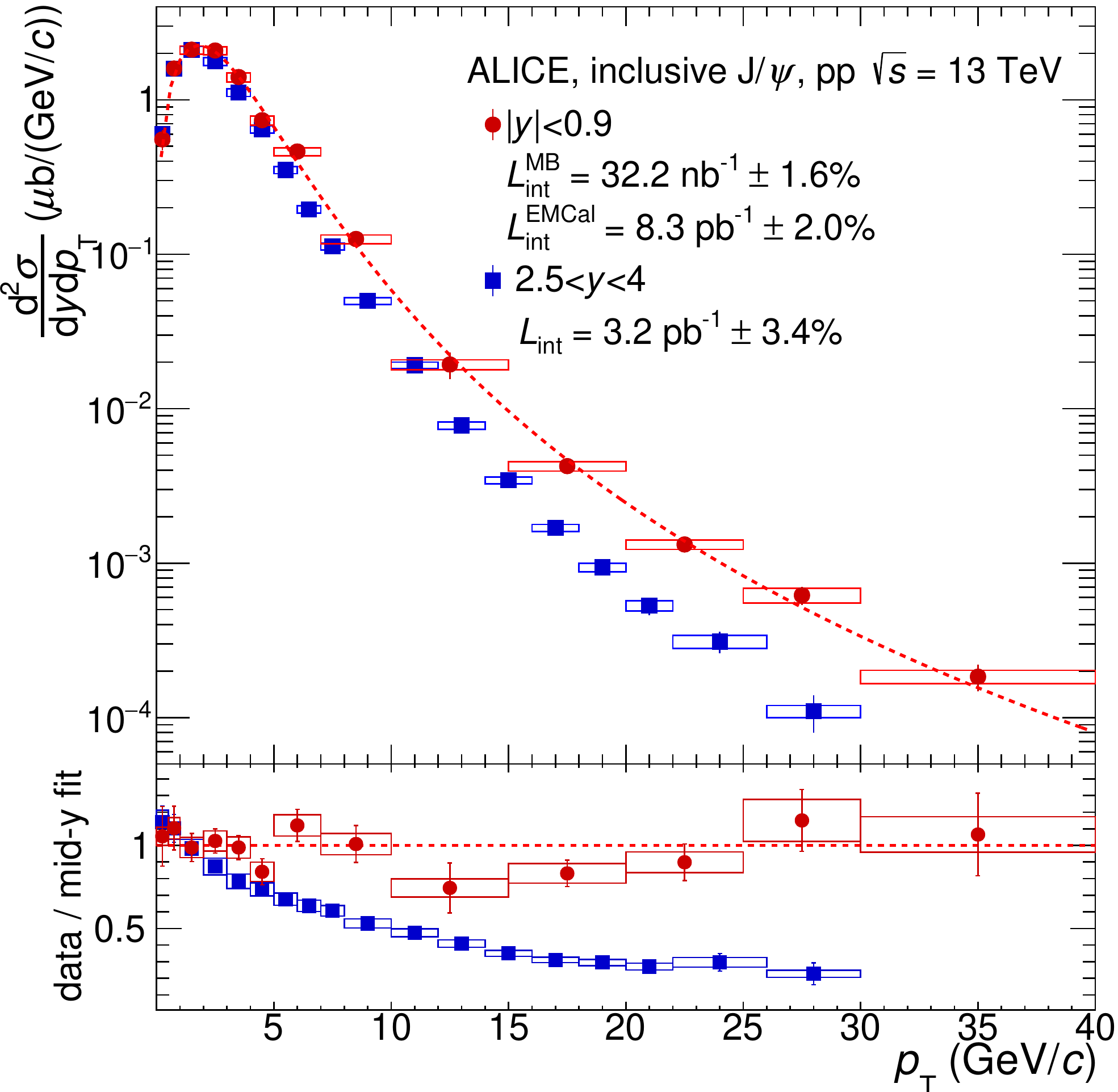}
    \includegraphics[width=.48\textwidth]{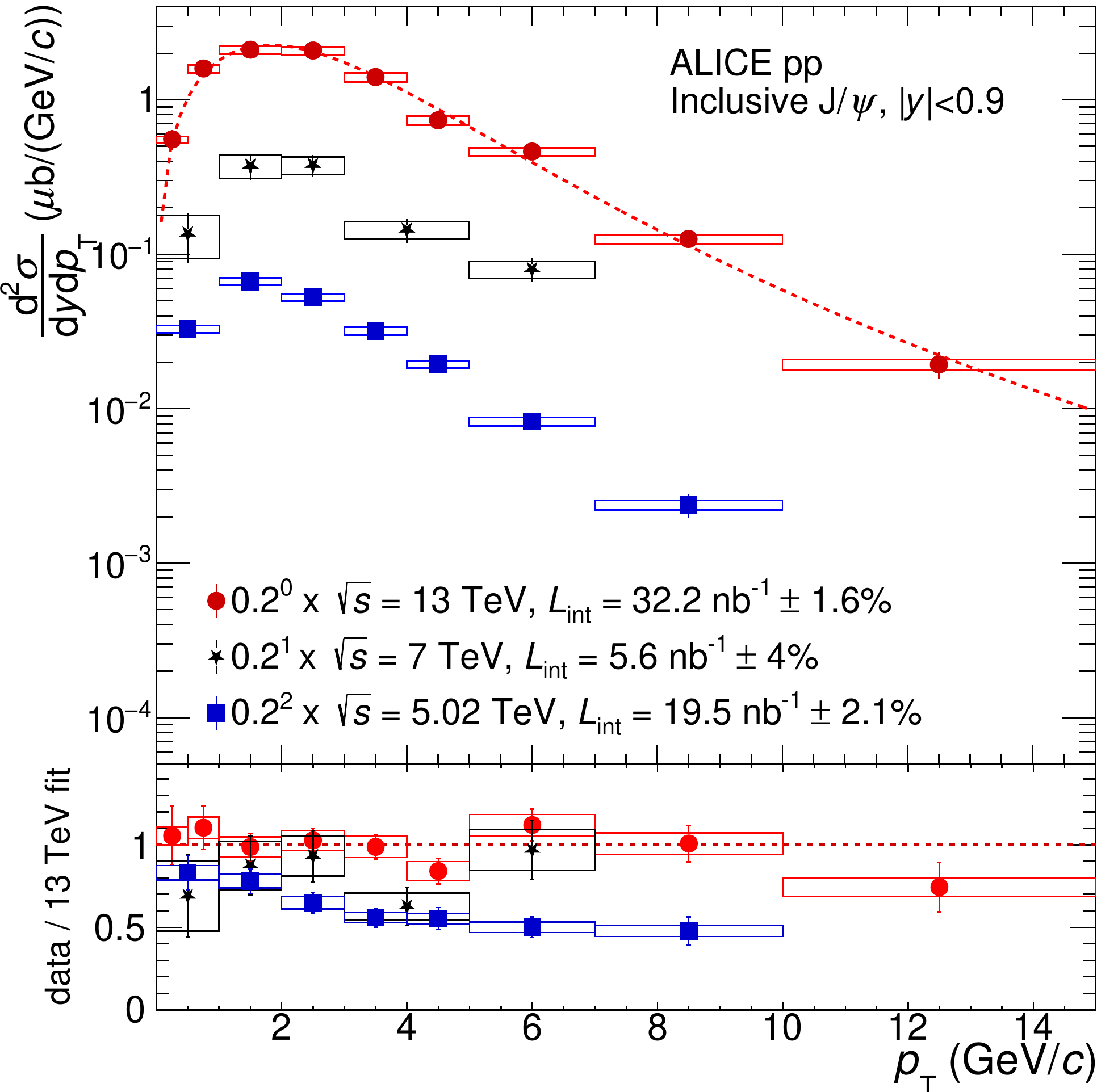}  
    \caption{Inclusive \jpsi production cross section at midrapidity ($|y|<0.9$) in \pp collisions at \thirteen compared with the ALICE forward-rapidity measurement at \thirteen~\cite{Acharya:2017hjh} (left panel) and the scaled ALICE midrapidity measurements at \five~\cite{Acharya:2019lkw} and \seven~\cite{Aamodt:2011gj} (right panel). The error bars represent statistical uncertainties while the boxes around the data points represent the total systematic uncertainty, excluding the global uncertainty from the luminosity and branching ratio. The lower panels show the ratio between the measurements at different rapidity and energies, and the power law fit, discussed in the text, to the \jpsi production cross section (red dashed line) at midrapidity at \thirteen. The boxes in the lower panels include the systematic uncertainty of the data points and the uncertainty of the integral of the fit function in the given \pt interval added in quadrature.} 
    \label{fig:crossSectionComparison}
\end{figure}

The inclusive \pt-differential \jpsi production cross section in \pp collisions at \thirteen at midrapidity, obtained from the combined analysis of the MB-triggered ($\pt<15$~\GeVc) and EMCal-triggered ($15<\pt<40$~\GeVc) samples is shown in the upper panels of Fig.~\ref{fig:crossSectionComparison}. Statistical uncertainties are shown as error bars, while the boxes around the data points represent the correlated and uncorrelated systematic uncertainties added in quadrature, excluding the global uncertainty from luminosity determination and branching ratio. The $x$-axis extent of the boxes illustrates the size of the \pt interval with the data points placed in the center. A simple power law function of the type 

\begin{equation}
    f(\pt) = A \times \pt / (1+(\pt/p_0)^2)^n,
    \label{eq:fitFunc}
\end{equation}

with $A$, $p_0$ and $n$ being free parameters, is used to fit the measured distribution. Since the systematic uncertainties are largely correlated, only the statistical ones are used in the fit. Due to the large \pt intervals, the fit is performed by considering the mean value of the function in the \pt interval rather than its value at the center of the interval. The values obtained for the fitted parameters are $A=2.15\pm0.18~\mu\text{b}/(\text{GeV}/c)^2$, $p_0=4.09\pm0.22$~\GeVc and $n=3.04\pm0.09$. The fit function, shown in Fig.~\ref{fig:crossSectionComparison} as a dashed red line, provides a good description of the data points measured with both MB and EMCal data samples, and illustrates the consistency between the two analyses.

In the left panel of Fig.~\ref{fig:crossSectionComparison}, the production cross section measured at midrapidity is compared with the forward rapidity measurement at the same energy~\cite{Acharya:2017hjh}. The bottom panel shows the ratio between the forward rapidity data points and the mean cross section at midrapidity obtained by integrating the fit function described above in the \pt intervals of the forward-rapidity measurement. The displayed uncertainty boxes include the systematic uncertainty of the forward rapidity measurement and the uncertainty of the function mean, added in quadrature. A monotonic drop of this ratio can be observed towards high \pt, indicating a harder \jpsi \pt distribution at midrapidity.

The right panel of Fig.~\ref{fig:crossSectionComparison} shows a comparison with midrapidity measurements performed by the ALICE Collaboration in \pp collisions at the lower collision energies of \seven~\cite{Aamodt:2011gj} and 5.02~TeV~~\cite{Acharya:2019lkw}. In the bottom panel, the ratio between the lower energy measurements and the fitted 13~TeV results is shown. The displayed uncertainty boxes include the systematic uncertainty of the lower energy data points and the uncertainty of the fit function mean, added in quadrature. Although the uncertainties of the measurement at 7~TeV are large, the data indicates an increase of the production cross section with increasing collision energy. In addition, the monotonic drop of the ratio between the 5.02~TeV and 13~TeV measurements indicates a hardening of the \pt spectrum with increasing collision energy, as also observed from the energy dependence of the inclusive \jpsi average \pt discussed in Ref.~\cite{Acharya:2019lkw}.   

\begin{figure}
    \centering
    \includegraphics[width=.48\textwidth]{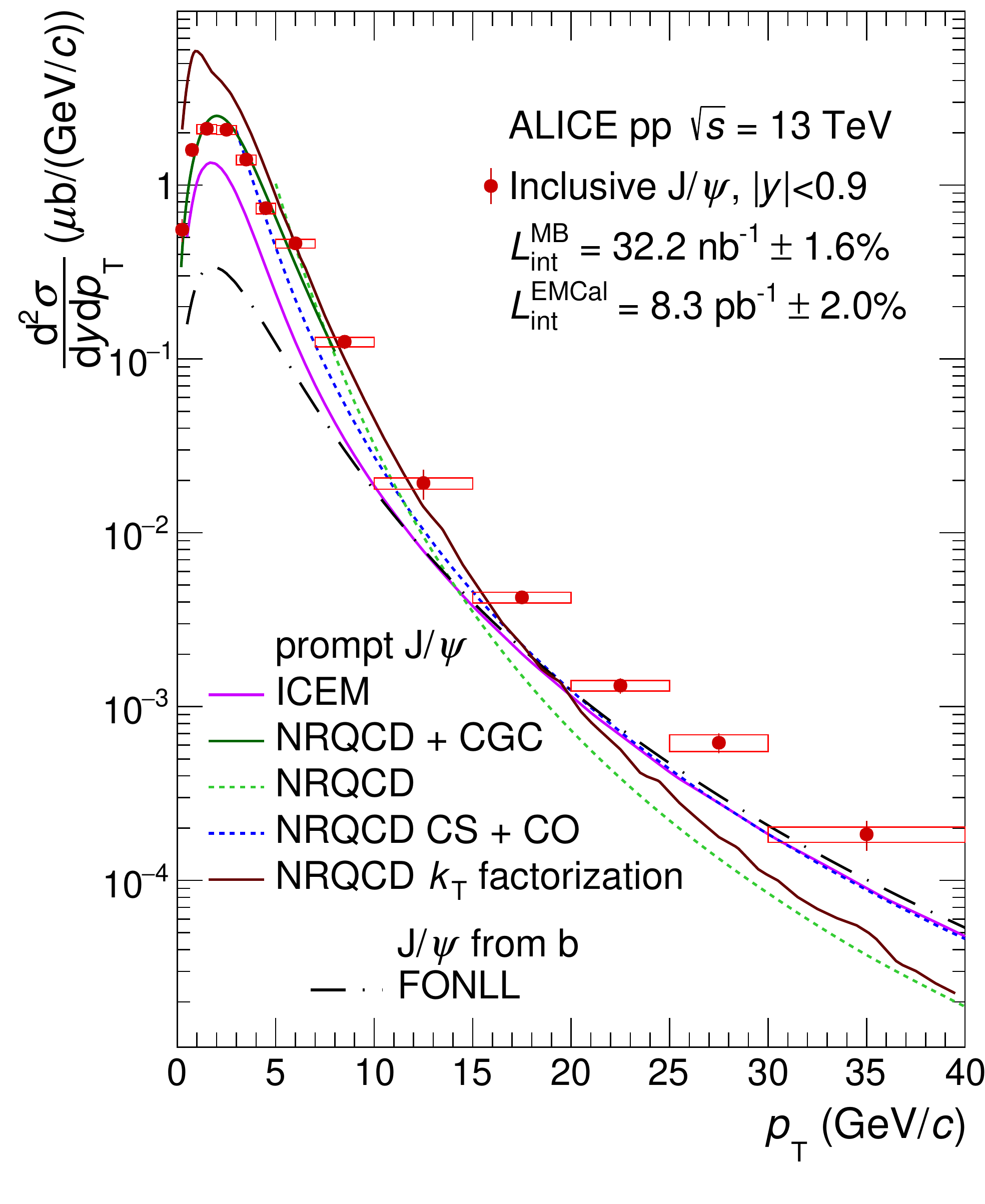}
    \includegraphics[width=.48\textwidth]{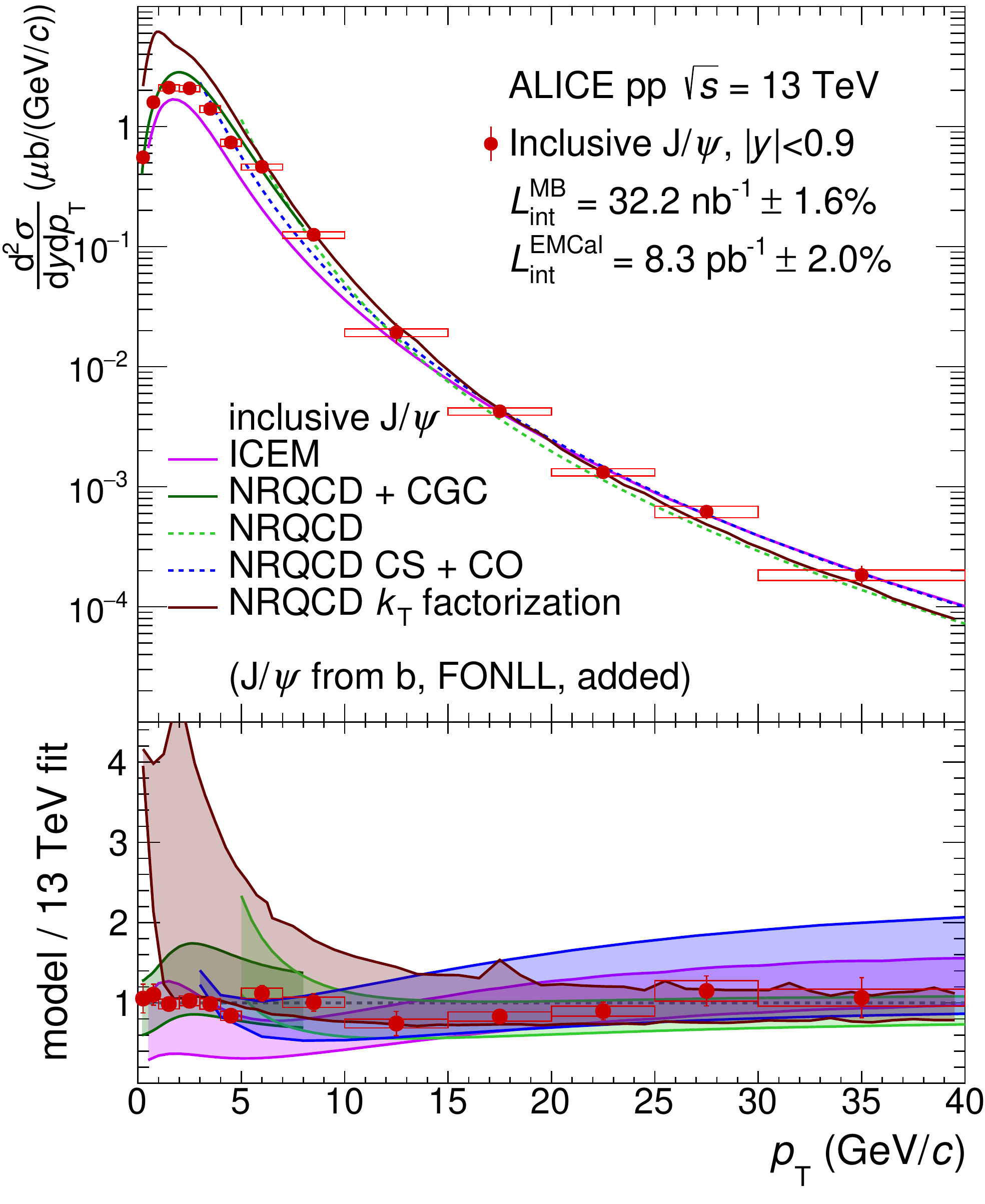}
    \caption{(Color online) Inclusive \jpsi production cross section compared with calculations for the prompt \jpsi production cross section using ICEM~\cite{Cheung:2018}, NLO NRQCD~\cite{Ma:2010yw,Butenschoen:2010rq,Lipatov:2019oxs}, LO NRQCD+CGC~\cite{Ma:2014mri} and for the non-prompt \jpsi from beauty-hadron feed-down using FONLL~\cite{Cacciari:2012ny} (left panel). Inclusive \jpsi production cross section compared with the corresponding calculations obtained as the sum of the prompt \jpsi component shown in the left panel and the non-prompt contribution from FONLL (right panel). The bottom panel shows the ratios between the model calculations and a fit to the data points. The bands illustrate the theoretical uncertainties centered around the ratio between the model calculation and the power-law fit to the data (see text for details).}
    \label{fig:differentialCrossSection}
\end{figure}

The measured inclusive \jpsi production cross section is compared with several phenomenological calculations of the prompt \jpsi production in the left panel of Fig.~\ref{fig:differentialCrossSection}. In addition, to illustrate the impact of the unaccounted feed-down from beauty decays in the theory predictions, a calculation of the non-prompt \jpsi production by Cacciari et al., using the Fixed-Order Next-to-Leading-Logarithms approach (FONLL)~\cite{Cacciari:2012ny}, is shown in the same panel. According to the FONLL calculations, the non-prompt contribution to the inclusive \jpsi yield is approximately 10\% at low-\pt and grows to approximately 50\% at \pt=40~\GeVc. In the right panel of Fig.~\ref{fig:differentialCrossSection}, the measured inclusive production cross section is compared to predictions for inclusive \jpsi production obtained as the sum of the prompt \jpsi calculations listed above and the beauty feed-down contribution calculated using FONLL. The bottom panel shows the ratio between each theoretical calculation and the fit to the data. The colored bands represent the theoretical uncertainties for each model, centered around the model to data ratio. These uncertainties are typically due to the variation of the renormalization and factorization scales, and of the charm quark mass.

Several phenomenological approaches are used for the calculation of the \jpsi yields shown in Fig.~\ref{fig:differentialCrossSection}. The green and blue dashed lines represent NLO NRQCD calculations from Ma et al.~\cite{Ma:2010yw} and Butenschoen et al.~\cite{Butenschoen:2010rq}, respectively, using collinear gluon parton distribution functions (PDFs). Although the calculation of the short distance terms is very similar, the predictions of these two approaches differ due to the LDME sets which are obtained in separate fits of the Tevatron and HERA data with different low-\pt cutoffs. In addition, the calculation from Ref.~\cite{Butenschoen:2010rq} does not include the feed-down from higher mass charmonium states. The brown solid line represents a calculation obtained with the MC generator PEGASUS~\cite{Lipatov:2019oxs} developed by Lipatov et al. which employs a \kt-factorization approach using \pt-dependent gluon distribution functions and NRQCD matrix elements combined with LDMEs extracted from an NLO high-transverse momentum analysis~\cite{Baranov:2019lhm}. Using the KMR~\cite{Kimber:2001sc} technique to construct the unintegrated gluon PDFs, this calculation can extend down to \jpsi $\pt=0$. A different model to calculate the low \pt \jpsi production cross section, by Ma and Venugopalan~\cite{Ma:2014mri} (green solid line) is based on a Color-Glass Condensate (CGC) approach coupled to a Leading Order (LO) NRQCD calculation which includes a soft-gluon resummation. The calculations obtained using the ICEM model by Cheung and Vogt~\cite{Cheung:2018} within the \kt-factorization approach are shown by the violet solid line. In this calculation, LDMEs are not used, however one normalization parameter per charmonium state is used to account for long distance effects~\cite{Nelson:2012bc}. The feed down from the higher mass charmonium states is taken into account in this model. 

As shown in the right panel of Fig.~\ref{fig:differentialCrossSection}, all the models provide a reasonable description of the inclusive \jpsi production cross section within the theoretical uncertainties over the entire \pt range covered by this measurement. In particular, both the ICEM and NRQCD+CGC calculations show very good agreement with the data in the low-\pt range. The NRQCD calculation from Lipatov, which uses the \kt-factorization approach, also provides a good description of the data for $\pt > 2$~\GeVc, while it overestimates the measured cross section at lower \pt. However, this is a significant progress compared to traditional collinear approaches which tend to diverge towards $\pt=0$.

\begin{figure}
    \centering
    \includegraphics[width=.48\textwidth]{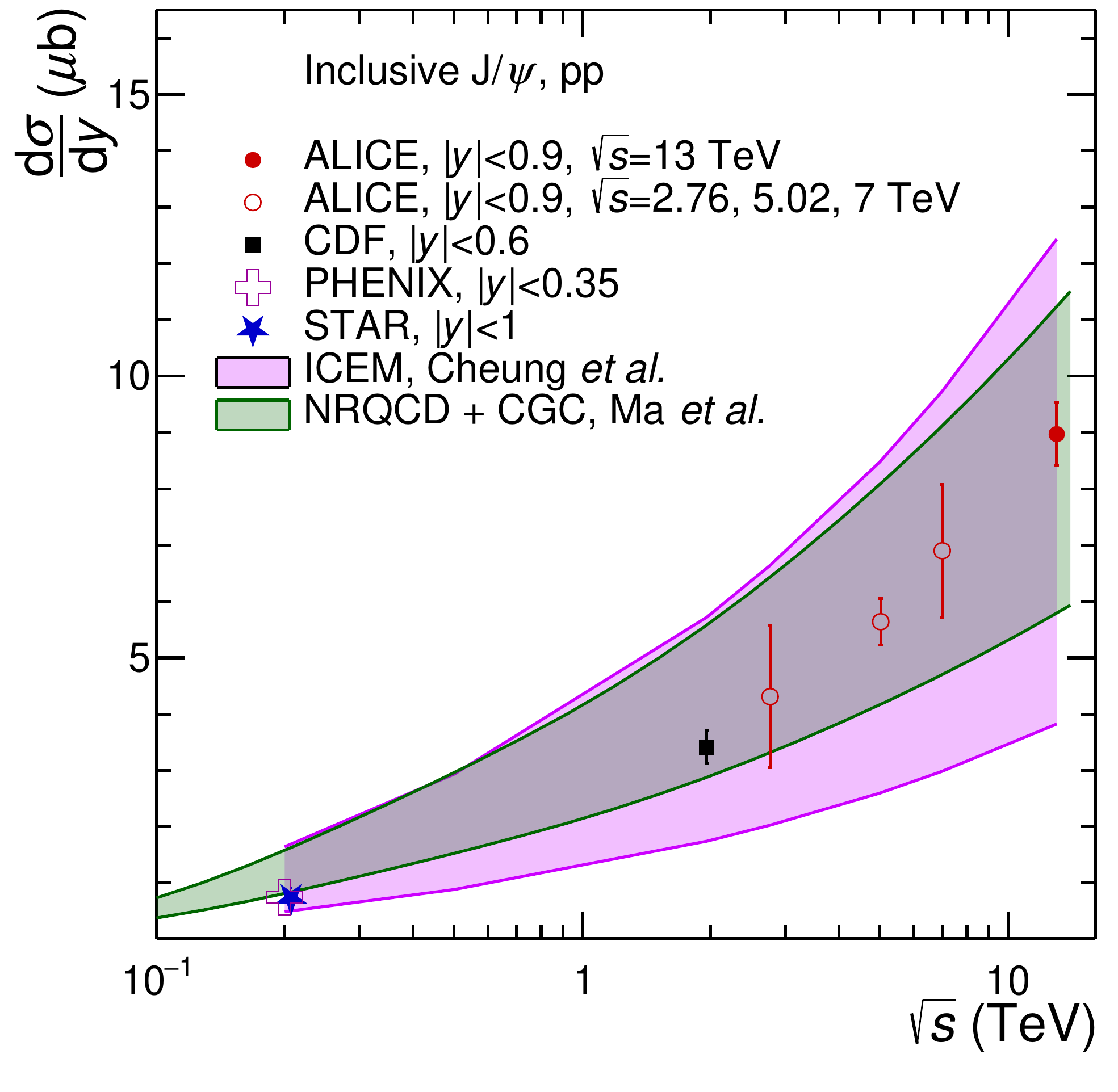}
    \includegraphics[width=.48\textwidth]{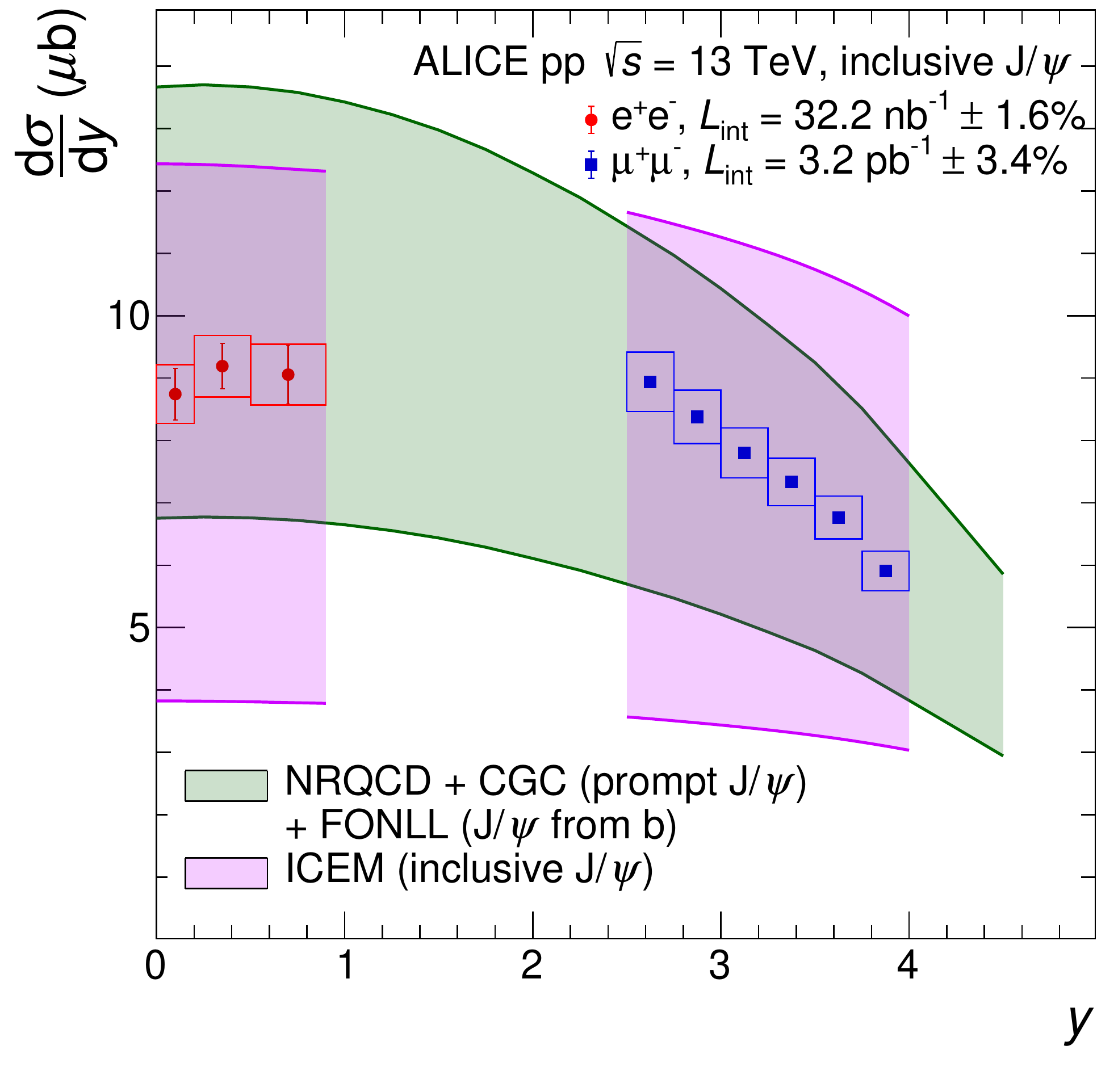}
    
    \caption{(Color online) Inclusive \pt-integrated \jpsi production cross section as a function of collision energy (left panel) and rapidity (right panel)
    compared with the ICEM~\cite{Cheung:2018} and NRQCD+CGC~\cite{Ma:2014mri} model calculations. The midrapidity \pt-integrated production cross section
    values are measured by the PHENIX~\cite{Adare:2011vq}, STAR~\cite{Adam:2018jmp}, CDF~\cite{Acosta:2004yw} and ALICE~\cite{Abelev:2012kr, Acharya:2019lkw, Aamodt:2011gj} collaborations. The forward-rapidity production cross section shown in the right panel is reported by ALICE in Ref.~\cite{Acharya:2017hjh}.
    }
    \label{fig:integratedCrossSection}
\end{figure}

The measured \pt-integrated inclusive \jpsi production cross section at midrapidity, obtained by performing the analysis on the MB data sample without any explicit selection on the \jpsi \pt ($\pt>0$), is  

\begin{align*}
    \frac{\text{d}\sigma_{\jpsi}}{\text{d}y} = 8.97 \pm 0.24 \; \text{(stat.)} \pm 0.48 \; \text{(syst.)} \pm 0.15 \; \text{(lumi)} \pm 0.05 \; \text{(BR)} \; \mu\text{b} \;.
\end{align*}

The systematic uncertainty includes all the systematic sources mentioned in Table~\ref{tab:systematics} added in quadrature, with the exception of the global ones which are given separately. The fraction of the total cross section covered by the EMCal-triggered event analysis results ($\pt>15$~\GeVc) is estimated to be less than 0.5\% and is not used in this measurement. A comparison of this measurement with previous midrapidity measurements in pp collisions at lower energies from PHENIX~\cite{Adare:2011vq}, STAR~\cite{Adam:2018jmp}, CDF~\cite{Acosta:2004yw}  and ALICE~\cite{Abelev:2012kr, Acharya:2019lkw, Aamodt:2011gj} is shown in the left panel of Fig.~\ref{fig:integratedCrossSection}. An approximate logarithmic increase of the cross sections with the energy is observed. The collision energy-dependent measurements are also compared with the calculations from the ICEM~\cite{Cheung:2018} and the NRQCD+CGC~\cite{Ma:2014mri} models. Both calculations provide a good description of the energy dependent trend within large theoretical uncertainties, dominated by the low-\pt region of the spectrum.

The right panel of Fig.~\ref{fig:integratedCrossSection} shows the rapidity-differential inclusive \jpsi production cross section which includes three data points measured in this analysis around midrapidity and the previously published ALICE measurement at forward rapidity~\cite{Acharya:2017hjh}. While no rapidity dependence is observed in the central rapidity range, a steep decrease towards forward rapidity is seen. The two model calculations employed, using ICEM~\cite{Cheung:2018} and the NRQCD+CGC~\cite{Ma:2014mri}, combined with non-prompt contributions calculated with FONLL~\cite{Cacciari:2012ny}, exhibit rather different rapidity dependences. However, both are compatible with the data owing to the large theoretical uncertainties, which are at present much larger than the experimental uncertainties.

\section{Conclusions}
\label{sec:conclusion}

The integrated, and the \pt- and $y$-differential inclusive \jpsi production cross sections at midrapidity ($|y|<0.9$) in \pp collisions at \thirteen are presented in the \pt range $0<\pt<40$~\GeVc, exceeding the \pt range of all the previous measurements reported by the ALICE Collaboration. The measurement up to 15~\GeVc is performed using a minimum-bias triggered data sample and the one at high-\pt is performed using an EMCal-triggered data sample, both collected by ALICE during the LHC Run 2.

The data are compared with the ALICE lower collision energy measurements at midrapidity~\cite{Acharya:2019lkw,Aamodt:2011gj} and with the ALICE forward-rapidity results~\cite{Acharya:2017hjh}. An approximate logarithmic dependence of the integrated \jpsi production cross section with collision energy is suggested by the data, in agreement with the available predictions. The \pt-differential cross section measured in this analysis shows a significant hardening with respect to both the forward-rapidity measurement at \thirteen and the midrapidity measurement at \five.

Several calculations within the NRQCD framework~\cite{Ma:2010yw,Butenschoen:2010rq,Lipatov:2019oxs,Ma:2014mri} and one using the ICEM approach~\cite{Cheung:2018} are compared with the measured inclusive \jpsi production cross section. In particular, the ICEM, the NRQCD model based on the CGC approach and a new NRQCD calculation using the \kt-factorization approach provide a prediction for the \jpsi production cross section down to $\pt = 0$. All models provide a good description of the measured inclusive \jpsi production cross section, although with large theoretical uncertainties.


\newenvironment{acknowledgement}{\relax}{\relax}
\begin{acknowledgement}
\section*{Acknowledgements}
We wish to thank Mathias Butenschoen, Vincent Cheung, Bernd A. Kniehl, Artem V. Lipatov, Yan-Qing Ma, Raju Venugopalan and Ramona Vogt for kindly providing their calculations.

The ALICE Collaboration would like to thank all its engineers and technicians for their invaluable contributions to the construction of the experiment and the CERN accelerator teams for the outstanding performance of the LHC complex.
The ALICE Collaboration gratefully acknowledges the resources and support provided by all Grid centres and the Worldwide LHC Computing Grid (WLCG) collaboration.
The ALICE Collaboration acknowledges the following funding agencies for their support in building and running the ALICE detector:
A. I. Alikhanyan National Science Laboratory (Yerevan Physics Institute) Foundation (ANSL), State Committee of Science and World Federation of Scientists (WFS), Armenia;
Austrian Academy of Sciences, Austrian Science Fund (FWF): [M 2467-N36] and Nationalstiftung f\"{u}r Forschung, Technologie und Entwicklung, Austria;
Ministry of Communications and High Technologies, National Nuclear Research Center, Azerbaijan;
Conselho Nacional de Desenvolvimento Cient\'{\i}fico e Tecnol\'{o}gico (CNPq), Financiadora de Estudos e Projetos (Finep), Funda\c{c}\~{a}o de Amparo \`{a} Pesquisa do Estado de S\~{a}o Paulo (FAPESP) and Universidade Federal do Rio Grande do Sul (UFRGS), Brazil;
Ministry of Education of China (MOEC) , Ministry of Science \& Technology of China (MSTC) and National Natural Science Foundation of China (NSFC), China;
Ministry of Science and Education and Croatian Science Foundation, Croatia;
Centro de Aplicaciones Tecnol\'{o}gicas y Desarrollo Nuclear (CEADEN), Cubaenerg\'{\i}a, Cuba;
Ministry of Education, Youth and Sports of the Czech Republic, Czech Republic;
The Danish Council for Independent Research | Natural Sciences, the VILLUM FONDEN and Danish National Research Foundation (DNRF), Denmark;
Helsinki Institute of Physics (HIP), Finland;
Commissariat \`{a} l'Energie Atomique (CEA) and Institut National de Physique Nucl\'{e}aire et de Physique des Particules (IN2P3) and Centre National de la Recherche Scientifique (CNRS), France;
Bundesministerium f\"{u}r Bildung und Forschung (BMBF) and GSI Helmholtzzentrum f\"{u}r Schwerionenforschung GmbH, Germany;
General Secretariat for Research and Technology, Ministry of Education, Research and Religions, Greece;
National Research, Development and Innovation Office, Hungary;
Department of Atomic Energy Government of India (DAE), Department of Science and Technology, Government of India (DST), University Grants Commission, Government of India (UGC) and Council of Scientific and Industrial Research (CSIR), India;
Indonesian Institute of Science, Indonesia;
Istituto Nazionale di Fisica Nucleare (INFN), Italy;
Institute for Innovative Science and Technology , Nagasaki Institute of Applied Science (IIST), Japanese Ministry of Education, Culture, Sports, Science and Technology (MEXT) and Japan Society for the Promotion of Science (JSPS) KAKENHI, Japan;
Consejo Nacional de Ciencia (CONACYT) y Tecnolog\'{i}a, through Fondo de Cooperaci\'{o}n Internacional en Ciencia y Tecnolog\'{i}a (FONCICYT) and Direcci\'{o}n General de Asuntos del Personal Academico (DGAPA), Mexico;
Nederlandse Organisatie voor Wetenschappelijk Onderzoek (NWO), Netherlands;
The Research Council of Norway, Norway;
Commission on Science and Technology for Sustainable Development in the South (COMSATS), Pakistan;
Pontificia Universidad Cat\'{o}lica del Per\'{u}, Peru;
Ministry of Education and Science, National Science Centre and WUT ID-UB, Poland;
Korea Institute of Science and Technology Information and National Research Foundation of Korea (NRF), Republic of Korea;
Ministry of Education and Scientific Research, Institute of Atomic Physics and Ministry of Research and Innovation and Institute of Atomic Physics, Romania;
Joint Institute for Nuclear Research (JINR), Ministry of Education and Science of the Russian Federation, National Research Centre Kurchatov Institute, Russian Science Foundation and Russian Foundation for Basic Research, Russia;
Ministry of Education, Science, Research and Sport of the Slovak Republic, Slovakia;
National Research Foundation of South Africa, South Africa;
Swedish Research Council (VR) and Knut \& Alice Wallenberg Foundation (KAW), Sweden;
European Organization for Nuclear Research, Switzerland;
Suranaree University of Technology (SUT), National Science and Technology Development Agency (NSDTA) and Office of the Higher Education Commission under NRU project of Thailand, Thailand;
Turkish Energy, Nuclear and Mineral Research Agency (TENMAK), Turkey;
National Academy of  Sciences of Ukraine, Ukraine;
Science and Technology Facilities Council (STFC), United Kingdom;
National Science Foundation of the United States of America (NSF) and United States Department of Energy, Office of Nuclear Physics (DOE NP), United States of America.
\end{acknowledgement}

\bibliographystyle{utphys}   
\bibliography{bibliography}

\providecommand{\href}[2]{#2}\begingroup\raggedright\begin{thebibliography}{10}

\bibitem{Brambilla:2010cs}
N.~Brambilla {\em et~al.}, ``{Heavy Quarkonium: Progress, Puzzles, and
  Opportunities}'',
  \href{http://dx.doi.org/10.1140/epjc/s10052-010-1534-9}{{\em Eur. Phys. J.}
  {\bfseries C71} (2011) 1534},
\href{http://arxiv.org/abs/1010.5827}{{\ttfamily arXiv:1010.5827 [hep-ph]}}.

\bibitem{Bodwin:1994jh}
G.~T. Bodwin, E.~Braaten, and G.~P. Lepage, ``{Rigorous QCD analysis of
  inclusive annihilation and production of heavy quarkonium}'',
  \href{http://dx.doi.org/10.1103/PhysRevD.55.5853,
  10.1103/PhysRevD.51.1125}{{\em Phys. Rev.} {\bfseries D51} (1995)
  1125--1171}, \href{http://arxiv.org/abs/hep-ph/9407339}{{\ttfamily
  arXiv:hep-ph/9407339 [hep-ph]}}.
[Erratum: Phys. Rev.D55,5853(1997)].

\bibitem{Butenschoen:2010rq}
M.~Butenschoen and B.~A. Kniehl, ``{Reconciling $J/\psi$ production at HERA,
  RHIC, Tevatron, and LHC with NRQCD factorization at next-to-leading order}'',
  \href{http://dx.doi.org/10.1103/PhysRevLett.106.022003}{{\em Phys. Rev.
  Lett.} {\bfseries 106} (2011) 022003},
\href{http://arxiv.org/abs/1009.5662}{{\ttfamily arXiv:1009.5662 [hep-ph]}}.

\bibitem{Ma:2010yw}
Y.-Q. Ma, K.~Wang, and K.-T. Chao, ``{$J/\psi (\psi^\prime)$ production at the
  Tevatron and LHC at ${\cal O}(\alpha_s^4v^4)$ in nonrelativistic QCD}'',
  \href{http://dx.doi.org/10.1103/PhysRevLett.106.042002}{{\em Phys. Rev.
  Lett.} {\bfseries 106} (2011) 042002},
\href{http://arxiv.org/abs/1009.3655}{{\ttfamily arXiv:1009.3655 [hep-ph]}}.

\bibitem{Chao:2012iv}
K.-T. Chao, Y.-Q. Ma, H.-S. Shao, K.~Wang, and Y.-J. Zhang, ``{$J/\psi$
  Polarization at Hadron Colliders in Nonrelativistic QCD}'',
  \href{http://dx.doi.org/10.1103/PhysRevLett.108.242004}{{\em Phys. Rev.
  Lett.} {\bfseries 108} (2012) 242004},
  \href{http://arxiv.org/abs/1201.2675}{{\ttfamily arXiv:1201.2675 [hep-ph]}}.

\bibitem{Butenschoen:2012px}
M.~Butenschoen and B.~A. Kniehl, ``{J/$\psi$ polarization at Tevatron and LHC:
  Nonrelativistic-QCD factorization at the crossroads}'',
  \href{http://dx.doi.org/10.1103/PhysRevLett.108.172002}{{\em Phys. Rev.
  Lett.} {\bfseries 108} (2012) 172002},
  \href{http://arxiv.org/abs/1201.1872}{{\ttfamily arXiv:1201.1872 [hep-ph]}}.

\bibitem{Baranov:2002cf}
S.~P. Baranov, ``Highlights from the ${k}_{T}$-factorization approach on the
  quarkonium production puzzles'',
  \href{http://dx.doi.org/10.1103/PhysRevD.66.114003}{{\em Phys. Rev. D}
  {\bfseries 66} (Dec, 2002) 114003}.
  \url{https://link.aps.org/doi/10.1103/PhysRevD.66.114003}.

\bibitem{Baranov:2019lhm}
S.~P. Baranov and A.~V. Lipatov, ``{Are there any challenges in the charmonia
  production and polarization at the LHC?}'',
  \href{http://dx.doi.org/10.1103/PhysRevD.100.114021}{{\em Phys. Rev. D}
  {\bfseries 100} no.~11, (2019) 114021},
  \href{http://arxiv.org/abs/1906.07182}{{\ttfamily arXiv:1906.07182
  [hep-ph]}}.

\bibitem{Kang:2013hta}
Z.-B. Kang, Y.-Q. Ma, and R.~Venugopalan, ``{Quarkonium production in high
  energy proton-nucleus collisions: CGC meets NRQCD}'',
  \href{http://dx.doi.org/10.1007/JHEP01(2014)056}{{\em JHEP} {\bfseries 01}
  (2014) 056}, \href{http://arxiv.org/abs/1309.7337}{{\ttfamily arXiv:1309.7337
  [hep-ph]}}.

\bibitem{Barger:1979js}
V.~D. Barger, W.-Y. Keung, and R.~Phillips, ``{On $\psi$ and $\Upsilon$
  production via gluons}'',
  \href{http://dx.doi.org/10.1016/0370-2693(80)90444-X}{{\em Phys. Lett. B}
  {\bfseries 91} (1980) 253--258}.

\bibitem{Gavai:1994in}
R.~Gavai, D.~Kharzeev, H.~Satz, G.~Schuler, K.~Sridhar, and R.~Vogt,
  ``{Quarkonium production in hadronic collisions}'',
  \href{http://dx.doi.org/10.1142/S0217751X95001443}{{\em Int. J. Mod. Phys. A}
  {\bfseries 10} (1995) 3043--3070},
  \href{http://arxiv.org/abs/hep-ph/9502270}{{\ttfamily arXiv:hep-ph/9502270}}.

\bibitem{Ma:2016exq}
Y.-Q. Ma and R.~Vogt, ``{Quarkonium Production in an Improved Color Evaporation
  Model}'', \href{http://dx.doi.org/10.1103/PhysRevD.94.114029}{{\em Phys. Rev.
  D} {\bfseries 94} no.~11, (2016) 114029},
  \href{http://arxiv.org/abs/1609.06042}{{\ttfamily arXiv:1609.06042
  [hep-ph]}}.

\bibitem{Chang:1979nn}
C.-H. Chang, ``{Hadronic Production of $J/\psi$ Associated With a Gluon}'',
  \href{http://dx.doi.org/10.1016/0550-3213(80)90175-3}{{\em Nucl. Phys. B}
  {\bfseries 172} (1980) 425--434}.

\bibitem{Baier:1981uk}
R.~Baier and R.~Ruckl, ``{Hadronic Production of J/$\psi$ and Upsilon:
  Transverse Momentum Distributions}'',
\href{http://dx.doi.org/10.1016/0370-2693(81)90636-5}{{\em Phys. Lett.}
  {\bfseries 102B} (1981) 364--370}.

\bibitem{Berger:1980ni}
E.~L. Berger and D.~L. Jones, ``{Inelastic Photoproduction of J/$\psi$ and
  Upsilon by Gluons}'', \href{http://dx.doi.org/10.1103/PhysRevD.23.1521}{{\em
  Phys. Rev. D} {\bfseries 23} (1981) 1521--1530}.

\bibitem{Lansberg:2010vq}
J.~Lansberg, ``{QCD corrections to J/$\psi$ polarisation in pp collisions at
  RHIC}'', \href{http://dx.doi.org/10.1016/j.physletb.2010.10.054}{{\em Phys.
  Lett. B} {\bfseries 695} (2011) 149--156},
  \href{http://arxiv.org/abs/1003.4319}{{\ttfamily arXiv:1003.4319 [hep-ph]}}.

\bibitem{Sirunyan:2017qdw}
{\bfseries CMS} Collaboration, A.~Sirunyan {\em et~al.}, ``{Measurement of
  quarkonium production cross sections in pp collisions at $\sqrt{s}=$ 13
  TeV}'', \href{http://dx.doi.org/10.1016/j.physletb.2018.02.033}{{\em Phys.
  Lett. B} {\bfseries 780} (2018) 251--272},
  \href{http://arxiv.org/abs/1710.11002}{{\ttfamily arXiv:1710.11002
  [hep-ex]}}.

\bibitem{Aamodt:2011gj}
{\bfseries ALICE} Collaboration, K.~Aamodt {\em et~al.}, ``{Rapidity and
  transverse momentum dependence of inclusive J$/\psi$ production in $pp$
  collisions at $\sqrt{s} = 7$ TeV}'',
  \href{http://dx.doi.org/10.1016/j.physletb.2011.09.054,
  10.1016/j.physletb.2012.10.060}{{\em Phys. Lett.} {\bfseries B704} (2011)
  442--455}, \href{http://arxiv.org/abs/1105.0380}{{\ttfamily arXiv:1105.0380
  [hep-ex]}}.
[Erratum: Phys. Lett.B718,692(2012)].

\bibitem{Abelev:2012kr}
{\bfseries ALICE} Collaboration, B.~Abelev {\em et~al.}, ``{Inclusive $J/\psi$
  production in $pp$ collisions at $\sqrt{s} = 2.76$ TeV}'',
  \href{http://dx.doi.org/10.1016/j.physletb.2012.10.078,
  10.1016/j.physletb.2015.06.058}{{\em Phys. Lett.} {\bfseries B718} (2012)
  295--306}, \href{http://arxiv.org/abs/1203.3641}{{\ttfamily arXiv:1203.3641
  [hep-ex]}}.
[Erratum: Phys. Lett.B748,472(2015)].

\bibitem{Acharya:2019lkw}
{\bfseries ALICE} Collaboration, S.~Acharya {\em et~al.}, ``{Inclusive J/$\psi$
  production at mid-rapidity in pp collisions at $\sqrt{s}$ = 5.02 TeV}'',
  \href{http://dx.doi.org/10.1007/JHEP10(2019)084}{{\em JHEP} {\bfseries 10}
  (2019) 084},
\href{http://arxiv.org/abs/1905.07211}{{\ttfamily arXiv:1905.07211 [nucl-ex]}}.

\bibitem{Acharya:2017hjh}
{\bfseries ALICE} Collaboration, S.~Acharya {\em et~al.}, ``{Energy dependence
  of forward-rapidity $\mathrm {J}/\psi $ and $\psi \mathrm {(2S)}$ production
  in pp collisions at the LHC}'',
  \href{http://dx.doi.org/10.1140/epjc/s10052-017-4940-4}{{\em Eur. Phys. J.}
  {\bfseries C77} no.~6, (2017) 392},
\href{http://arxiv.org/abs/1702.00557}{{\ttfamily arXiv:1702.00557 [hep-ex]}}.

\bibitem{Aaij:2015rla}
{\bfseries LHCb} Collaboration, R.~Aaij {\em et~al.}, ``{Measurement of forward
  $J/\psi$ production cross-sections in $pp$ collisions at $\sqrt{s}=13$
  TeV}'', \href{http://dx.doi.org/10.1007/JHEP10(2015)172}{{\em JHEP}
  {\bfseries 10} (2015) 172}, \href{http://arxiv.org/abs/1509.00771}{{\ttfamily
  arXiv:1509.00771 [hep-ex]}}. [Erratum: JHEP 05, 063 (2017)].

\bibitem{Aamodt:2008zz}
{\bfseries ALICE} Collaboration, K.~Aamodt {\em et~al.}, ``{The ALICE
  experiment at the CERN LHC}'',
\href{http://dx.doi.org/10.1088/1748-0221/3/08/S08002}{{\em JINST} {\bfseries
  3} (2008) S08002}.

\bibitem{Abelev:2014ffa}
{\bfseries ALICE} Collaboration, B.~B. Abelev {\em et~al.}, ``{Performance of
  the ALICE Experiment at the CERN LHC}'',
  \href{http://dx.doi.org/10.1142/S0217751X14300440}{{\em Int. J. Mod. Phys.}
  {\bfseries A29} (2014) 1430044},
\href{http://arxiv.org/abs/1402.4476}{{\ttfamily arXiv:1402.4476 [nucl-ex]}}.

\bibitem{Aamodt:2010aa}
{\bfseries ALICE} Collaboration, K.~Aamodt {\em et~al.}, ``{Alignment of the
  ALICE Inner Tracking System with cosmic-ray tracks}'',
  \href{http://dx.doi.org/10.1088/1748-0221/5/03/P03003}{{\em JINST} {\bfseries
  5} (2010) P03003},
\href{http://arxiv.org/abs/1001.0502}{{\ttfamily arXiv:1001.0502
  [physics.ins-det]}}.

\bibitem{Alme:2010ke}
J.~Alme {\em et~al.}, ``{The ALICE TPC, a large 3-dimensional tracking device
  with fast readout for ultra-high multiplicity events}'',
  \href{http://dx.doi.org/10.1016/j.nima.2010.04.042}{{\em Nucl. Instrum.
  Meth.} {\bfseries A622} (2010) 316--367},
\href{http://arxiv.org/abs/1001.1950}{{\ttfamily arXiv:1001.1950
  [physics.ins-det]}}.

\bibitem{Abeysekara:2010ze}
{\bfseries ALICE EMCal} Collaboration, U.~Abeysekara {\em et~al.}, ``{ALICE
  EMCal Physics Performance Report}'',
  \href{http://arxiv.org/abs/1008.0413}{{\ttfamily arXiv:1008.0413
  [physics.ins-det]}}.

\bibitem{Cortese:2008zza}
{\bfseries ALICE} Collaboration, P.~Cortese {\em et~al.}, ``{ALICE
  electromagnetic calorimeter technical design report}'', Tech. Rep.
  CERN-LHCC-2008-014, CERN-ALICE-TDR-014, 9, 2008.

\bibitem{Allen:2010stl}
J.~Allen {\em et~al.}, ``{ALICE DCal: An Addendum to the EMCal Technical Design
  Report Di-Jet and Hadron-Jet correlation measurements in ALICE}'', Tech. Rep.
  CERN-LHCC-2010-011, ALICE-TDR-14-add-1, 6, 2010.

\bibitem{Abbas:2013taa}
{\bfseries ALICE} Collaboration, E.~Abbas {\em et~al.}, ``{Performance of the
  ALICE VZERO system}'',
  \href{http://dx.doi.org/10.1088/1748-0221/8/10/P10016}{{\em JINST} {\bfseries
  8} (2013) P10016},
\href{http://arxiv.org/abs/1306.3130}{{\ttfamily arXiv:1306.3130 [nucl-ex]}}.

\bibitem{vanderMeer:296752}
S.~van~der Meer, ``{Calibration of the effective beam height in the ISR}'',
  Tech. Rep. CERN-ISR-PO-68-31. ISR-PO-68-31, CERN, Geneva, 1968.
\newblock \url{http://cds.cern.ch/record/296752}.

\bibitem{aliceLumi13TeVrun2}
{\bfseries ALICE} Collaboration, S.~Acharya {\em et~al.}, ``{ALICE
  2016-2017-2018 luminosity determination for pp collisions at $\sqrt{s}=$
  13TeV}'', Tech. Rep. ALICE-PUBLIC-2021-005, CERN, 2021.
\newblock \url{https://cds.cern.ch/record/2776672/}.

\bibitem{Acharya:2018kkj}
{\bfseries ALICE} Collaboration, S.~Acharya {\em et~al.}, ``{Dielectron and
  heavy-quark production in inelastic and high-multiplicity
  proton\textendash{}proton collisions at $\sqrt {s_{NN}}=$ 13TeV}'',
  \href{http://dx.doi.org/10.1016/j.physletb.2018.11.009}{{\em Phys. Lett. B}
  {\bfseries 788} (2019) 505--518},
  \href{http://arxiv.org/abs/1805.04407}{{\ttfamily arXiv:1805.04407
  [hep-ex]}}.

\bibitem{Tanabashi:2018oca}
{\bfseries Particle Data Group} Collaboration, M.~Tanabashi {\em et~al.},
  ``{Review of Particle Physics}'',
\href{http://dx.doi.org/10.1103/PhysRevD.98.030001}{{\em Phys. Rev.} {\bfseries
  D98} no.~3, (2018) 030001}.

\bibitem{Sjostrand:2006za}
T.~Sjostrand, S.~Mrenna, and P.~Z. Skands, ``{PYTHIA 6.4 Physics and Manual}'',
  \href{http://dx.doi.org/10.1088/1126-6708/2006/05/026}{{\em JHEP} {\bfseries
  05} (2006) 026},
\href{http://arxiv.org/abs/hep-ph/0603175}{{\ttfamily arXiv:hep-ph/0603175
  [hep-ph]}}.

\bibitem{Skands:2010ak}
P.~Z. Skands, ``{Tuning Monte Carlo Generators: The Perugia Tunes}'',
  \href{http://dx.doi.org/10.1103/PhysRevD.82.074018}{{\em Phys. Rev. D}
  {\bfseries 82} (2010) 074018},
  \href{http://arxiv.org/abs/1005.3457}{{\ttfamily arXiv:1005.3457 [hep-ph]}}.

\bibitem{Bossu:2011qe}
F.~Bossu, Z.~C. del Valle, A.~de~Falco, M.~Gagliardi, S.~Grigoryan, and
  G.~Martinez~Garcia, ``{Phenomenological interpolation of the inclusive
  J/$\psi$ cross section to proton-proton collisions at 2.76 TeV and 5.5
  TeV}'',
\href{http://arxiv.org/abs/1103.2394}{{\ttfamily arXiv:1103.2394 [nucl-ex]}}.

\bibitem{Barberio:1990ms}
E.~Barberio, B.~van Eijk, and Z.~Was, ``{PHOTOS: A Universal Monte Carlo for
  QED radiative corrections in decays}'',
\href{http://dx.doi.org/10.1016/0010-4655(91)90012-A}{{\em Comput. Phys.
  Commun.} {\bfseries 66} (1991) 115--128}.

\bibitem{Brun:1082634}
R.~Brun, F.~Bruyant, F.~Carminati, S.~Giani, M.~Maire, A.~McPherson,
  G.~Patrick, and L.~Urban,
  \href{http://dx.doi.org/10.17181/CERN.MUHF.DMJ1}{{\em {GEANT: Detector
  Description and Simulation Tool; Oct 1994}}}.
\newblock CERN Program Library. CERN, Geneva, 1993.
\newblock \url{https://cds.cern.ch/record/1082634}.
\newblock Long Writeup W5013.

\bibitem{Sjostrand:2007gs}
T.~Sj{\"o}strand, S.~Mrenna, and P.~Z. Skands, ``{A Brief Introduction to
  PYTHIA 8.1}'', \href{http://dx.doi.org/10.1016/j.cpc.2008.01.036}{{\em
  Comput. Phys. Commun.} {\bfseries 178} (2008) 852--867},
\href{http://arxiv.org/abs/0710.3820}{{\ttfamily arXiv:0710.3820 [hep-ph]}}.

\bibitem{Acharya:2017jgo}
{\bfseries ALICE} Collaboration, S.~Acharya {\em et~al.}, ``{Measurement of
  D-meson production at mid-rapidity in pp collisions at ${\sqrt{s}=7}$ TeV}'',
  \href{http://dx.doi.org/10.1140/epjc/s10052-017-5090-4}{{\em Eur. Phys. J.}
  {\bfseries C77} no.~8, (2017) 550},
\href{http://arxiv.org/abs/1702.00766}{{\ttfamily arXiv:1702.00766 [hep-ex]}}.

\bibitem{Cheung:2018}
V.~Cheung and R.~Vogt, ``{Production and polarization of prompt
  $J/\ensuremath{\psi}$ in the improved color evaporation model using the
  ${k}_{T}$-factorization approach}'',
  \href{http://dx.doi.org/10.1103/PhysRevD.98.114029}{{\em Phys. Rev. D}
  {\bfseries 98} (Dec, 2018) 114029}.
  \url{https://link.aps.org/doi/10.1103/PhysRevD.98.114029}.

\bibitem{Lipatov:2019oxs}
A.~Lipatov, M.~Malyshev, and S.~Baranov, ``{Particle Event Generator: A
  Simple-in-Use System PEGASUS version 1.0}'',
  \href{http://dx.doi.org/10.1140/epjc/s10052-020-7898-6}{{\em Eur. Phys. J. C}
  {\bfseries 80} no.~4, (2020) 330},
  \href{http://arxiv.org/abs/1912.04204}{{\ttfamily arXiv:1912.04204
  [hep-ph]}}.

\bibitem{Ma:2014mri}
Y.-Q. Ma and R.~Venugopalan, ``{Comprehensive Description of J/$\psi$
  Production in Proton-Proton Collisions at Collider Energies}'',
  \href{http://dx.doi.org/10.1103/PhysRevLett.113.192301}{{\em Phys. Rev.
  Lett.} {\bfseries 113} no.~19, (2014) 192301},
\href{http://arxiv.org/abs/1408.4075}{{\ttfamily arXiv:1408.4075 [hep-ph]}}.

\bibitem{Cacciari:2012ny}
M.~Cacciari, S.~Frixione, N.~Houdeau, M.~L. Mangano, P.~Nason, and G.~Ridolfi,
  ``{Theoretical predictions for charm and bottom production at the LHC}'',
  \href{http://dx.doi.org/10.1007/JHEP10(2012)137}{{\em JHEP} {\bfseries 10}
  (2012) 137},
\href{http://arxiv.org/abs/1205.6344}{{\ttfamily arXiv:1205.6344 [hep-ph]}}.

\bibitem{Kimber:2001sc}
M.~Kimber, A.~D. Martin, and M.~Ryskin, ``{Unintegrated parton
  distributions}'', \href{http://dx.doi.org/10.1103/PhysRevD.63.114027}{{\em
  Phys. Rev. D} {\bfseries 63} (2001) 114027},
  \href{http://arxiv.org/abs/hep-ph/0101348}{{\ttfamily arXiv:hep-ph/0101348}}.

\bibitem{Nelson:2012bc}
R.~Nelson, R.~Vogt, and A.~Frawley, ``{Narrowing the uncertainty on the total
  charm cross section and its effect on the J/$\psi$ cross section}'',
  \href{http://dx.doi.org/10.1103/PhysRevC.87.014908}{{\em Phys. Rev. C}
  {\bfseries 87} no.~1, (2013) 014908},
  \href{http://arxiv.org/abs/1210.4610}{{\ttfamily arXiv:1210.4610 [hep-ph]}}.

\bibitem{Adare:2011vq}
{\bfseries PHENIX} Collaboration, A.~Adare {\em et~al.}, ``{Ground and excited
  charmonium state production in $p+p$ collisions at $\sqrt{s}=200$ GeV}'',
  \href{http://dx.doi.org/10.1103/PhysRevD.85.092004}{{\em Phys. Rev.}
  {\bfseries D85} (2012) 092004},
\href{http://arxiv.org/abs/1105.1966}{{\ttfamily arXiv:1105.1966 [hep-ex]}}.

\bibitem{Adam:2018jmp}
{\bfseries STAR} Collaboration, J.~Adam {\em et~al.}, ``{$J/\psi$ production
  cross section and its dependence on charged-particle multiplicity in $p + p$
  collisions at $\sqrt{s}$ = 200 GeV}'',
  \href{http://dx.doi.org/10.1016/j.physletb.2018.09.029}{{\em Phys. Lett.}
  {\bfseries B786} (2018) 87--93},
\href{http://arxiv.org/abs/1805.03745}{{\ttfamily arXiv:1805.03745 [hep-ex]}}.

\bibitem{Acosta:2004yw}
{\bfseries CDF} Collaboration, D.~Acosta {\em et~al.}, ``{Measurement of the
  $J/\psi$ meson and $b-$hadron production cross sections in $p\bar{p}$
  collisions at $\sqrt{s} = 1960$ GeV}'',
  \href{http://dx.doi.org/10.1103/PhysRevD.71.032001}{{\em Phys. Rev.}
  {\bfseries D71} (2005) 032001},
\href{http://arxiv.org/abs/hep-ex/0412071}{{\ttfamily arXiv:hep-ex/0412071
  [hep-ex]}}.

\end{thebibliography}\endgroup

\newpage
\appendix
\section{The ALICE collaboration}
\label{app:collab}

\begin{flushleft}

S.~Acharya$^{\rm 143}$, 
D.~Adamov\'{a}$^{\rm 98}$, 
A.~Adler$^{\rm 76}$, 
G.~Aglieri Rinella$^{\rm 35}$, 
M.~Agnello$^{\rm 31}$, 
N.~Agrawal$^{\rm 55}$, 
Z.~Ahammed$^{\rm 143}$, 
S.~Ahmad$^{\rm 16}$, 
S.U.~Ahn$^{\rm 78}$, 
I.~Ahuja$^{\rm 39}$
% ...

\bigskip

\bigskip 

\textbf{\Large Affiliation Notes}

\bigskip 

% ...
$^{\rm VII}$ Also at: University of Kansas, Lawrence, Kansas, United States\\

\bigskip

\bigskip 

\textbf{\Large Collaboration Institutes}

\bigskip 

% ...
$^{16}$ Department of Physics, Aligarh Muslim University, Aligarh, India\\
% ...
$^{31}$ Dipartimento DISAT del Politecnico and Sezione INFN, Turin, Italy\\
% ...
$^{35}$ European Organization for Nuclear Research (CERN), Geneva, Switzerland\\
% ...
$^{39}$ Faculty of Science, P.J.~\v{S}af\'{a}rik University, Ko\v{s}ice, Slovakia\\
% ...
$^{55}$ INFN, Sezione di Bologna, Bologna, Italy\\
% ...
$^{76}$ Johann-Wolfgang-Goethe Universit\"{a}t Frankfurt Institut f\"{u}r Informatik, Fachbereich Informatik und Mathematik, Frankfurt, Germany\\
% ...
$^{78}$ Korea Institute of Science and Technology Information, Daejeon, Republic of Korea\\
% ...
$^{98}$ Nuclear Physics Institute of the Czech Academy of Sciences, \v{R}e\v{z} u Prahy, Czech Republic\\
% ...
$^{143}$ Variable Energy Cyclotron Centre, Homi Bhabha National Institute, Kolkata, India\\
% ...

\bigskip 

\end{flushleft} 
  
\end{document}